\documentclass[journal]{IEEEtran}
\usepackage{cite}
\usepackage{bm}
\usepackage{amsmath}
\usepackage{extarrows}
\usepackage{amssymb}
\usepackage{graphicx}
\usepackage{xcolor}
\usepackage{enumerate}
\usepackage{bookmark} 
\graphicspath{{figure}}
\usepackage{stfloats}
\usepackage{setspace}
\usepackage{multirow}
\usepackage{amsmath} 
\usepackage{framed} 
\usepackage{esint}
\usepackage{subfigure}

\newcommand{\mr}{\mathrm}

\newcommand{\BE}{\begin{equation}}
\newcommand{\EE}{\end{equation}}
\newcommand{\BS}{\begin{subequations}}
\newcommand{\ES}{\end{subequations}}
\renewcommand{\bf}{\bm}

\renewcommand{\cal}{\mathcal}

\newtheorem{theorem}{Theorem}

\newtheorem{assumption}[theorem]{Assumption}
\newtheorem{definition}[theorem]{Definition}

\newtheorem{remark}[theorem]{Remark}
\newtheorem{lemma}[theorem]{Lemma}

\newtheorem{example}[theorem]{Example}

\ifCLASSINFOpdf
\graphicspath{{figure/}}
\else
\fi
\begin{document}
\title{\huge Orthogonal AMP for Problems with Multiple Measurement Vectors and/or Multiple Transforms}
\author{{ Yiyao~Cheng, Lei~Liu, \emph{Member, IEEE}, Shansuo~Liang, \\Jonathan~H.~Manton, \emph{Fellow, IEEE}, and~Li~Ping, \emph{Fellow, IEEE}}

\thanks{Yiyao~Cheng and~Li~Ping are with the Department of Electrical Engineering, CityU, Hong Kong, SAR, China (e-mail: yiycheng2-c@my.cityu.edu.hk, eeliping@cityu.edu.hk).}
\thanks{Lei~Liu was with the Department of Electronic Engineering, City University of Hong Kong (CityU), Hong Kong, SAR, China, and is currently with the School of Information Science, Japan Institute of Science and Technology (JAIST), Ishikawa 923-1292, Japan (e-mail: leiliu@jaist.ac.jp).}
\thanks{Shansuo Liang is with the Theory Lab, Central Research Institute, 2012 Labs, Huawei Technologies Co. Ltd, Hong Kong, SAR, China (e-mail:liang.shansuo@huawei.com).}
\thanks{Jonathan~H.~Manton is with the Department of Electrical and Electronic Engineering, The University of Melbourne, VIC 3010, Australia (e-mail: j.manton@ieee.org).}
}

\maketitle

\begin{abstract}
Approximate message passing (AMP) algorithms break a (high-dimensional) statistical problem into parts then repeatedly solve each part in turn, akin to alternating projections. A distinguishing feature is their asymptotic behaviours can be accurately predicted via their associated state evolution equations.
Orthogonal AMP (OAMP) was recently developed to avoid the need for computing the so-called Onsager term in traditional AMP algorithms, providing two clear benefits: the derivation of an OAMP algorithm is both straightforward and more broadly applicable. OAMP was originally demonstrated for statistical problems with a single measurement vector and single transform.
This paper extends OAMP to statistical problems with multiple measurement vectors (MMVs) and multiple transforms (MTs). We name the resulting algorithms as OAMP-MMV and OAMP-MT respectively, and their combination as augmented OAMP (A-OAMP). Whereas the extension of traditional AMP algorithms to such problems would be challenging, the orthogonal principle underpinning OAMP makes these extensions straightforward.

The MMV and MT models are widely applicable to signal processing and communications. We present an example of MIMO relay system with correlated source data and signal clipping, which can be modelled as a joint MMV-MT system. While existing methods meet with difficulties in this example, OAMP offers an efficient solution with excellent performance.



\end{abstract}


\IEEEpeerreviewmaketitle
\vspace{0.5cm}

\section{Introduction}
\subsection{Problem Formulation}
\begin{figure} 
\centering
 \subfigure[A linear transform system with two constraints $\Phi$ and $\Gamma$.]{
\begin{minipage}[b]{0.38\textwidth} 
\includegraphics[width=\textwidth]{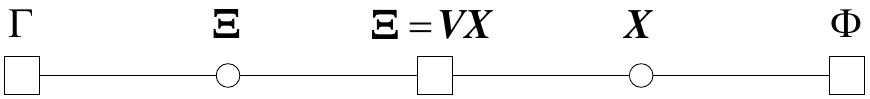} 
\end{minipage}   
}
\subfigure[Generic iterative processing (GIP) for the system in \eqref{Eqn:X_Vx_MMV}]{
\begin{minipage}[b]{0.37\textwidth} 
\includegraphics[width=\textwidth]{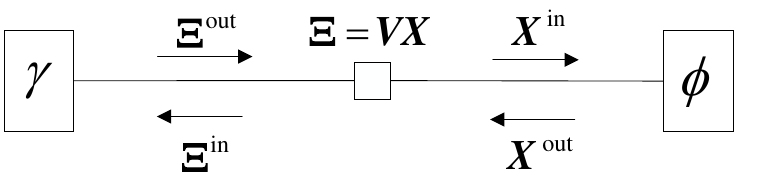} 
\end{minipage} 
}
\caption{Graphical illustration for an MMV system}\label{Fig:MMV_net}
\end{figure}

Fig. \ref{Fig:MMV_net}(a) illustrates a problem involving two variables ${\bm \Xi}$ and ${\bm X}$ connected by a transform:
\begin{equation} \label{Eqn:MMV_MC}
{\bm{ \Xi}} = {\bm V}{\bm X},
\end{equation}
where ${\bm V}$ is an orthogonal matrix. Two constraints $\Gamma$ and $\Phi$ are applied to ${\bm \Xi}$ and ${\bm X}$ respectively. Our task is to estimate ${\bm X}$ (or equivalently ${\bm \Xi}$). For simplicity, we assume that ${\bm \Xi}$, ${\bm V}$ and ${\bm X}$ are all real in this paper unless otherwise stated. The results can be extended to the
complex case straightforwardly.

For example, consider a linear model:
\begin{equation}\label{Eqn:linear_system}
\bf{Y}=\bf{SX}+\bf{N},
\end{equation}
where the columns of ${\bm Y}$ are called measurement vectors, $\bf{S}$ a sensing matrix and $\bf{N}$ a noise matrix. The columns of ${\bm X}$ are vectors to be estimated. Assume that the entries of ${\bm X}$ are sampled from a finite alphabet. Such an example arises in communication systems where the alphabet is given by a modulation constellation. For convenience, we will express such a constraint as an {\textit{a priori}} distribution $p(\bf{
X})$. 

Let the singular value decomposition (SVD) of $\bf{S}$ be $\bf{S}=\bf{U}\bf{\Lambda V}$ where ${\bm \Lambda }$ is diagonal, and ${\bm U}$ and ${\bm V}$ are both orthogonal. Define $\bf{R}\equiv\bf{U}^{\mr{T}}\bf{Y}$, $\bf{\Upsilon}\equiv\bf{U}^{\mr{T}}\bf{N}$ and $\bf{\Xi}\equiv\bf{VX}$. Then \eqref{Eqn:linear_system} can be represented by Fig. \ref{Fig:MMV_net}(a) using the following two constraints:
\vspace{-0.25cm}\BS\label{Eqn:linear_sys2}\begin{align}
&\Gamma: \quad  {\bf{R}}={\bm \Lambda }\bf{\Xi} + \bf{\Upsilon},\label{Eqn:linear_sys2a}\\
&\Phi: \quad {\bm X}\sim p(\bf{X}).\label{Eqn:linear_sys2b}
\end{align}\ES	
Clearly, \eqref{Eqn:linear_sys2} can be represented by Fig. \ref{Fig:MMV_net}(a).

\underline{SMV and MMV Problems:} If ${\bm X}$ contains only one column, we call Fig. \ref{Fig:MMV_net}(a) a single measurement vector (SMV) problem. Otherwise, we call it a multiple measurement vector (MMV) one since typically multiple measurements are involved, each providing information about a particular column of ${\bm X}$. If there is a correlation between different columns of ${\bm X}$, an MMV problem can be substantially more difficult than an SMV one.

\underline{ST and MT Problems:} The problem in Fig. \ref{Fig:MMV_net}(a) involves a single transform (ST). A more general one may involve multiple transforms (MTs), say ${\bm V}_1$, ${\bm V}_2,\cdots,{\bm V}_K$. Two MT examples are shown in Fig. \ref{Fig:MC_network}(a) and Fig. \ref{Fig:example1} for $K\!=\!2$ and $K\!=\!4$ respectively (which will be detailed in Sec. IV). In practice, the transforms $\{{\bm V}_k\}$ can be related to (but not limited to) the transforms to different domains such as time, frequency and space.

MMV and MT problems are found in a wide range of applications, e.g.,  direction of arrival (DOA) estimation \cite{DOA2015,DOA2018}, magnetic resonance imaging (MRI) \cite{MRI2008,MRI2009}, multiple-input multiple-output (MIMO) systems \cite{MIMO2018,MIMO2015} and cognitive radio networks \cite{CRN2010}. An example of an MIMO-relay system will be given in Sec. V below.

\subsection{Approximate Message Passing (AMP)}
Optimal estimations in the above problems using all the available pieces of statistical information are generally prohibitively complicated. Approximate message passing (AMP) is a low-cost technique for such problems. AMP is an iterative process converging to a parameter estimate by cyclically applying each piece of statistical information in turn, akin to alternating projections. This involves local estimators in each domain plus a global combining/exchanging process of the outputs of different local processors. Data incest is a main problem for such an iterative process. AMP uses an Onsager term, an idea borrowed from statistical mechanics, to avoid this problem \cite{Donoho2009}. 

AMP was originally devised for the SMV-ST model in \eqref{Eqn:linear_system} with ${\bm S}$ comprising of independently and identically distributed Gaussian (IID-Gaussian or IIDG) entries \cite{Donoho2009}. A distinguishing feature of AMP is a state evolution (SE) technique for predicting the mean-square error (MSE) at each iteration with proven asymptotic accuracy \cite{Bayati2011}, which provides a probabilistic performance assurance for AMP. In principle, one can seek to optimize the performance by altering the estimates used at each step, guided by SE. 
It has been widely reported that AMP outperforms other alternatives, such as the celebrated turbo algorithm \cite{Wang1999}, in many signal processing and communication applications. It has also been observed that AMP may perform poorly when ${\bm S}$ is not IID. 

AMP has been extended to MMV-ST and SMV-MT problems separately \cite{Yuwei2018,Hannak2018,Kim2011,Manoel2017}, but, to the best of our knowledge, there are restrictions in the available AMP schemes for MMV or MT problems. For example, similar to the standard AMP, IIDG sensing matrices are assumed in AMP-based MMV algorithms\cite{Yuwei2018, Kim2011}. The method studied in \cite{Manoel2017} is for a specific cascading MT structure with alternative linear and non-linear constraints. The SE for the existing AMP-based MMV and MT problems are either proved on a case-by-case basis \cite{Manoel2017} or verified empirically \cite{Yuwei2018}. Overall, there is still a lack of a unified framework for solving MMV and MT problems using AMP.

\subsection{Orthogonal AMP (OAMP)}
Orthogonal AMP, introduced in \cite{Ma2016} and elaborated on \cite{Lei_TSP_I,Ma2019,Yiyao2021}, was also originaly devised for the SMV-ST model in \eqref{Eqn:linear_system}. OAMP can work with unitarily-invariant sensing matrices \cite{Ma2016,Ma2019,Yiyao2021}. IIDG matrices are unitarily-invariant, so OAMP has a broader application range than AMP. 

OAMP \cite{Ma2016} comprises two aspects. First, instead of the Onsager term, OAMP employs an orthogonal principle to suppress data incest during iterative processing. This is a principle rather than an algorithm, because the user is largely free to choose the detailed updating functions (i.e., the prototype local estimators) at each step. 

The second aspect of OAMP is a constructive method for a working algorithm with the required orthogonality. This is first discussed in \cite{Ma2016} using a differential approach and later in \cite{Yiyao2021} using an integral approach. The latter employs a Gram-schmidt orthogonalization (GSO) technique and is more robust than the former. 

The SE analysis of OAMP in \cite{Takeuchi2020} provides a probabilistic performance assurance for OAMP. The optimality of OAMP is studied under the minimum MSE (MMSE) measure in \cite{Ma2016} and also under the mutual information measure in \cite{Lei2021}. Various applications of OAMP have been reported for signal processing and communications problems \cite{Ma2019,He2020,ZhangOAMP2017,Zhang2019,Xue2016,Yuan2022}.

\subsection{Contributions of This Paper}
In this paper, we investigate OAMP-based solutions for systems beyond SMV and ST. We define error orthogonality in MMV and MT environments and derive GSO techniques to realize such orthogonality. This leads to two algorithms for MMV and MT problems respectively, namely OAMP-MMV and OAMP-MT. These two algorithms can be combined to handle problems involving both MMVs and MTs. We refer to the resulted algorithm as augmented OAMP (A-OAMP). (This is a briefer name than a somewhat cumbersome alternative “OAMP-MMV-MT”).

We analyze the error behaviour of OAMP in the MMV and MT cases. The problem boils down to tracking error statistics when combining multiple messages. Here a message is a variable that carriers information about ${\bm X}$ (or ${\bm X}_1, \cdots,{\bm X}_K$ in the MT case), such as a noisy observation of ${\bm X}$, or an estimate of ${\bm X}$ that may contain errors. Combining multiple messages may hopefully generate an improved estimate for ${\bm X}$, but this is not guaranteed unless we know the statistical dependency among the messages to be combined. To overcome the difficulty of tracking such dependency, we borrow a technique from \cite{Bayati2011}: we assess the performance averaged over independently and uniformly sampled $\{{\bm V}_k\}$. For such average performance, we show that the orthogonality in A-OAMP suppresses the so-called error angles (see Sec. \ref{Sec:Error_OAMPMMV} below) at the input and output of a local estimator. Consequently, we only need to track the dependency of the norms of errors for A-OAMP. This is a relatively easy task due to the greatly reduced problem dimensions. SE is a basically recursive process for such tracking. 

The discussions related to SE in this paper in spirit follow  \cite{Bayati2011,Takeuchi2020,Takeuchi,Rangan2016,Dudeja2022}. The related derivations in Sec. IV-D, Sec.~IV-E and Appendix are more concise than those in \cite{Bayati2011,Takeuchi2020,Takeuchi,Rangan2016,Dudeja2022}, even though the MMV and MT problems studied in this paper are much more complex. This clearly shows the advantage of the orthogonal framework underpinning OAMP. On the other hand, the discussions in this paper are intuitive rather than rigorous. Our main aim is to reveal useful insights into the mechanism of A-OAMP in MMV and MT problems.

We demonstrate the application of A-OAMP using an MIMO-relay example with multiple correlated source data streams. Due to the use of MIMO, the received signal at the relay node has a high dynamical range. A clipping technique is used at the relay to improve amplifier power efficiency. We will model such a system as a joint MMV and MT problem. To the best of our knowledge, so far, there is no efficient treatment for this problem. We will show that A-OAMP can significantly outperform the crude method of ignoring clipping. Incidentally, the correlated source model may find other applications such as efficient large-scale population screening techniques for pandemic control \cite{Iru2021}. We will briefly discuss this possibility. 


Overall, we will show that, under the orthogonal principle, it is straightforward to extend the findings for OAMP from the basic SMV-ST model to more complicated MMV and MT cases. It opens promising directions for the future research into other applications. 



\subsection{Notation}
Boldface lowercase letters represent column vectors and boldface uppercase symbols denote matrices. $\bf{I}^{N \times N}$ denotes an $N \times N$ identity matrix, $\bf{0}^{N\times M}$ the $N\times M$ zero matrix, ${\bm {\mathcal{U}}}^{N \times N}$ the set of all $N\times N$ unitary matrices, and ${\bm {\mathcal{H}}}^{N \times N}$ the $N$-dimensional Haar distribution. For given matrices ${\bf{A}}, {\bf B} \in {{\mathbb{R}}^{N \times M}}$, ${\rm Sp}({\bf{A}})$ denotes the column space of $\bf A$, ${\rm Sp}({\bf{A}})^\perp$ denotes the null space of ${\rm Sp}({\bf{A}})$, and ${\rm Sp}({\bf{A}}) \perp {\rm Sp}({\bf{B}})$ means that ${\rm Sp}({\bf{A}})$ and ${\rm Sp}({\bf{B}})$ are mutually orthogonal or equivalently ${{\bm A}^{\rm{T}}}{\bm B} = {{\bm 0}^{M \times M}}$. ${\rm{Rank}}(\bf A)$ denotes the rank of $\bf A$, ${\rm{Clmn}}(\bf A)$ the column set of $\bf A$ and $\mathop\mr{ACF}\limits_{\mathcal{R}}\big\{\bf{A}\big\} \equiv \tfrac{1}{N}\mathop  \mr{E}\limits_{\mathcal{R}}\big\{\bf{A}^{\mr{T}}\bf{A}\big \}$ the ``auto-covariance'' function over a set of random variables $\mathcal{R}$. 

If we say that a matrix $\bf A$ is IID, we mean that its entry set $\{ {{A_{i,j}}} \}$ is IID. Similar sayings are used for other distributions such as IIDG or pair-wise IIDG (PIIDG) that will be discussed in Sec.~\ref{sec:preliminaries}.

\section{Preliminaries}
\label{sec:preliminaries}
\newcommand{\R}{\mathbb{R}}
\newcommand{\bX}{\bm{\mathcal{X}}}
\subsection{Notations for Matrix Functions}
Let $\bm G$ and $\bm F$ be two size $N\times J$ matrices connected by a function
\BE \label{Eqn:matrix_function}
{\bm G}=\psi(\bm F).
\EE
Assume that $J=K\times M$. We use the following notations.
\begin{itemize}
\item ${\bm G}_{-n}$ denotes the $n^{\rm{th}}$ row of ${\bm G}$.
\item ${\bm G}_{|j}$ denotes the $j^{\rm{th}}$ column of ${\bm G}$.
\item ${\bm G}_{||k}$ denotes the $k^{\rm{th}}$ column-wise block of ${\bm G}$, i.e., 
\BE \label{Eqn:partion}
{{\bm G}_{||k}} = [ {{\bm G}_{|(k - 1)M + 1}, {{\bm G}_{|(k - 1)M + 2}}, \cdots ,{{\bm G}_{|kM}}} ].
\EE
\end{itemize}
Using \eqref{Eqn:partion}, we can partition ${\bm G}$ into $K$ blocks as follows.
\BE \label{Eqn:partion_K}
\begin{aligned}
G=\big[\underbrace{{{\bm G}_{|1}}, \cdots ,{{\bm G}_{|M}}}_{{\bm G}_{||1}},\; \cdots\; ,\underbrace{{{\bm G}_{|(k - 1)M + 1}}, \cdots ,{{\bm G}_{|kM}}}_{{\bm G}_{||k}}, \\
\;\cdots\; ,\underbrace{{{\bm G}_{|(K - 1)M + 1}}, \;\cdots\; ,{{\bm G}_{|KM}}}_{{\bm G}_{||K}}\big].
\end{aligned}
\EE
Similar notations apply to ${\bm F}$.

We introduce a row-wise function
\BE \label{Eqn:row-wise}
{\bm G}_{-n}=\psi_{-n}({\bm F}_{-n}).
\EE
The above notation is nominal since, from \eqref{Eqn:matrix_function}, ${\bm G}_{-n}$ is a function of all rows of ${\bm F}$. In other words, we treat $\{{\bm F}_{-n'}, n'=1,2,\cdots,N, n'\ne n\}$ as implicit parameters in $\psi_{-n}(\cdot)$ in \eqref{Eqn:row-wise}.

Similarly, we introduce a column block-wise function
\BE \label{Eqn:block-wise}
{\bm G}_{||k}=\psi_{||k}({\bm F}_{||k}).
\EE
We call the pair $({\bm G}_{||k},{\bm F}_{||k})$ the $k^{\rm{th}}$ port of $\psi(\cdot)$ and call ${\bm G}_{||k}\!=\!\psi_{||k}({\bm F}_{||k})$ the $k^{\rm{th}}$ port-wise function of $\psi(\cdot)$. Clearly, similar to \eqref{Eqn:row-wise}, $\{{\bm F}_{||k}, k'\!=\!1,2,\!\cdots\!,K, k' \ne k\}$ are also implicit parameters of $\psi_{||k}(\cdot)$. We say that ${\bm G}=\psi(\bm F)$ is of single-port and multi-port respectively when $K=1$ and $K>1$. 

\subsection{Column-wise IID Matrices}
\begin{definition}[CIID Matrix] \label{Def:CIID}
 A random matrix $\bm{A} \in \R^{N \times M}$ is column-wise IID (CIID) if every column of ${\bm A}$ is IID.
\end{definition}

The following is easy to verify.
\begin{lemma} \label{Lem:CIID_1} 
Let ${\bm A}$ be CIID. Then all rows of ${\bm A}$ have the same auto-covariance, i.e., ${\rm{E}}\{ {{\bm A}_{ - n}^{\rm{T}}{{\bm A}_{ - n}}} \} = {\rm{E}}\{ {{\bm A}_{ - 1}^{\rm{T}}{{\bm A}_{ - 1}}} \}$ for $n=1,2,\cdots,N$. 
\end{lemma}

\subsection{Row-wise Separable Functions}
\begin{definition}[Separable and IID-separable Functions]  \label{Def:IID-separable}
Let ${\bm G } = \psi ( {\bm F} )$ and ${\bm G}_{-n}=\psi_{-n}({\bm F}_{-n})$.
\begin{itemize}
\item $\psi(\cdot)$ is (row-wise) separable if ${\psi}_{-n}(\cdot)$ is not a function of ${\bm F}_{-n'}, \forall n' \ne n$.
\item $\psi(\cdot)$ is IID-separable if (a) $\psi(\cdot)$ is (row-wise) separable, and (b) ${\bm G}$ is CIID provided that ${\bm F}$ is.
\end{itemize}
\end{definition}

We will only consider row-wise separability. Therefore, for brevity, ``separable'' implies ``row-wise separable'' in this paper. 

\begin{example}
Let $\psi(\cdot)$ be separable with $\psi_{-n}(\cdot)$ IID drawn from an ensemble $\mathbb{S}_{\psi}$ of functions. Then $\psi(\cdot)$ is IID-separable. In particular when $\psi_{-n}(\cdot)=\psi^{*}(\cdot),\forall n$, i.e., $\mathbb{S}_{\psi}$ contains only one element $\psi^{*}(\cdot)$. \end{example}

\begin{example}
Let $\psi(\cdot)$ be given in a row-by-row form as
\BE
\!\!{{\bm G} _{-n}} = {\psi _{-n}}( {{{\bm F}_{-n}}} )\! =\! {{\bm F}_{-n}}{\bm P} + {{\bm Q} _{-n}}, n=1,2,\cdots,N,
\EE
where ${\bm P}$ and ${\bm Q}_{-n}$ are respectively of sizes $J\times J$ and $1\times J$. Then $\psi(\cdot)$ is separable. Furthermore, $\psi(\cdot)$ is IID-separable if $\{{\bm P}\}$ and $\{{\bm Q}_{-n}\}$ are both CIID. However, $\psi(\cdot)$ is separable but not IID separable if either $\{{\bm P}\}$ or $\{{\bm Q}_{-n}\}$ is not CIID. 
\end{example}

The CIID property is preserved when a variable passes an IID-separable function. This makes it easy in tracking an iterative process, as we will see later. 
\subsection{Multi-Port Function and Separability}
Now assume that $K>1$ in \eqref{Eqn:matrix_function} and so ${\bm G}=\psi(\bm F)$ is a multi-port function. Let $\psi(\cdot)$ be IID-separable. In this case, since $\{{\bm F}_{||k'}, k'\ne k\}$ are parameters in ${\bm G}_{||k}=\psi_{||k}({\bm F}_{||k})$ (see Sec. II-A), $\psi_{||k}({\bm F}_{||k})$ may not be IID-separable unless all $\{{\bm F}_{||k'}, k'\ne k\}$ are also CIID. 
\begin{lemma} \label{Lem:multi-port-IID}
Let ${\bm G}=\psi({\bm F})$. The port-wise function ${\bm G}_{||k}=\psi_{||k}({\bm F}_{||k})$ is IID-separable when (i) $\psi(\cdot)$ is IID-separable and (ii) ${\bm F}$ is CIID.
\end{lemma}

We emphasise that requirement (ii) in Lemma~\ref{Lem:multi-port-IID} is that ${\bm F}$ is CIID, not just ${\bm F}_{||k}$. This is important for the discussions in Sec.~IV.

\subsection{The Haar Distribution and PIIDG Variables}
Let ${ \bm {\mathcal{U}}^{N \times N}}$ be the set of all $N\times N$ unitary matrices. The random matrix $\bf V$ is Haar distributed if its distribution is uniform over ${\bm {\mathcal{U}}}^{N \times N}$, denoted as ${\bf V} \sim {{\bm {\mathcal H}}^{N \times N}}$.
\begin{lemma} [Haar] \label{lem:Haar}
 Let $\bf V \sim {{\bm{\mathcal H}}^{N \times N}}$ be Haar distributed. Then, for any given ${\bf f }\in {{\mathbb R}^N}$ with the $l_2$-norm $\| {\bf f} \| > 0$, ${\bf V}{\bf f}$ follows a uniformly spherical distribution (USD) on an $N$-dimensional sphere of radius $\| {\bf f} \|$ centred at the origin. Consequently, for any ${\bm \Omega}\in \bm{\cal{U}}^{N \times N}$,  ${\bm \Omega} {\bm V}\sim \bm{\cal{H}}^{N\times N}$ and ${\bm V}{\bm \Omega}\sim \bm{\cal{H}}^{N \times N}$.
\end{lemma}
\begin{IEEEproof}
See Theorem 3.7 in \cite{Mattila1995}. 
\end{IEEEproof}

\begin{definition}[PIID and PIIDG Vector] \label{Def:PIIDG}
A set $\{ {{x_n}} \}$ is pair-wise IID if any two elements in $\{x_n\}$ are mutually IID. Specially, $\{x_n\}$ are PIID Gaussian (PIIDG) if $\{x_n\}$ are PIID as well as Gaussian. We say that a vector ${\bm f}$ or a matrix ${\bm F}$ is PIID or PIIDG if its entries are PIID or PIIDG respectively.
\end{definition}

\begin{lemma} \label{Lem:generalV}
Let ${\bm V}\sim \bm{\cal{H}}^{N \times N}$ conditional on ${\bm f} \in \mathbb{R}^{N \times 1}$. When $N\to \infty$, we have (i) ${\bm V}{\bm f}$ converges to PIIDG in distribution; and (ii)  ${\bm V}{\bm f}/\| \bm f \|$ is independent of ${\bm f}$ and $\bm z$ for $\bm z$ in a Markov chain ${\bm z} \to {\bm f} \to {\bm {Vf}}$.
\end{lemma}
\begin{IEEEproof}
(i) follows Theorem 2.8 in \cite{Meckes2014}. From Lemma~\ref{lem:Haar}, ${\bm V}{\bm f}/\| \bm f \|$ is USD on a $N$-dimensional unit sphere of radius of 1, which is independent of $\bm f$ and $\bm z$, so (ii) holds.
\end{IEEEproof}

Note that ${\bm {Vf}}$ is asymptotically PIIDG but not fully IID since $\| \bm {Vf} \| = \| \bm f \|$.
\subsection{Column-wise PIID (CPIID) and PIID Separable}
We can define a CPIID matrix and a PIID-separable function by changing IID to PIID in Definitions~\ref{Def:CIID} and \ref{Def:IID-separable} respectively. Then Lemmas~\ref{Lem:CIID_1} and \ref{Lem:multi-port-IID} can be modified accordingly to Lemmas~\ref{Lem:CPIID_1} and \ref{Lem:multi-port-PIID} below. We omit details since they are straightforward. 

\begin{lemma} \label{Lem:CPIID_1} 
Let ${\bm A}$ be CPIID. Then all rows of ${\bm A}$ have the same auto-covariance, i.e., ${\rm{E}}\{ {{\bm A}_{ - n}^{\rm{T}}{{\bm A}_{ - n}}} \} = {\rm{E}}\{ {{\bm A}_{ - 1}^{\rm{T}}{{\bm A}_{ - 1}}} \}$ for $n=1,2,\cdots,N$. 
\end{lemma}
\begin{lemma} \label{Lem:multi-port-PIID}
Let ${\bm G}=\psi({\bm F})$. The port-wise function ${\bm G}_{||k}=\psi_{||k}({\bm F}_{||k})$ is PIID-separable when (i) $\psi(\cdot)$ is PIID-separable and (ii) ${\bm F}$ is CPIID.
\end{lemma}

Incidentally, the following lemma is also straightforward. 
\begin{lemma} \label{Lem:CPIID_2}
${\bm A}$ is CPIID if all $\{{\bm A}_{||k}\}$ are CPIID, where $\{{\bm A}_{||k}\}$ are the blocks defined in \eqref{Eqn:partion_K}.
\end{lemma}

\subsection{Orthogonality under the Law of Large Numbers (LLN)} \label{Sec:LLNorthogonality}
\begin{definition}
 Let $\bf{f}=\{f_{n}\}$ and $\bf{g}=\{g_{n}\}$ be two sequences of length $N$, $\mr{E}\{\|\bf{f}\|^2\}\neq 0$,  $\mr{E}\{\|\bf{g}\|^2\}\neq 0$, $ \mr{E}\{f_n g_n\}=0$ and  $ \mr{Var}\{ f_n g_n\}$ finite. We say that the vectors ${\bm f}$ and $\bm g$ are LLN-orthogonal (see \cite{Lei_TSP_I}) denoted as  $\tfrac{1}{N}{{\bm f}^{\rm{T}}}{\bm g}\stackrel{\text{LLN}}{\longrightarrow}0$ if for any fixed $\delta>0$ and $\varepsilon>0$, there is a fixed $N'$ such that
\begin{equation} \label{Eqn:LLN}
\Pr \left(\! {\frac{{| {{{\bm f}^{\rm T}}{\bm g}} |}}{{\sqrt {{\rm{E}}\{ {\| {\bm f} \|_2^2} \}{\rm{E}}\{ {\| {\bm g} \|_2^2} \}} }} < \varepsilon } \!\right) \!\ge\! 1 - \delta \text{ for }N>N'.
\end{equation}
\end{definition}

LLN-orthogonality is stronger than the common orthogonality requirement ${\rm{E}}\{ {{{\bm f}^{\rm{T}}}{\bm g}} \} = 0$. The former ensures $\tfrac{1}{N}{{\bm f}^{\rm{T}}}{\bm g} \to { 0}$ for every experiment in probability when $N \to \infty $, while the latter does not. 

Using the law of large numbers, we can show that $\bm f$ and $\bm g$ are LLN-orthogonal provided that $\{{f_n}{g_n}\}$ are IID or PIID and ${\rm{E}}\{ {{{\bm f}^{\rm{T}}}{\bm g}} \} = 0$. Hence the name ''LLN-orthogonality''. We omit details due to space limitations. This holds when $\bm g$ is a function of $\bm f$, so we have the following.

\begin{lemma}\label{Lem:LLN}
Assume that ${\bm g }= \psi ( {\bm f })$ is PIID-separable and ${\bm f}$ is PIID. Let $N \to \infty$. Then $\tfrac{1}{N}{{\bm f}^{\rm{T}}} {\bm g} \stackrel{\text{LLN}}{\longrightarrow} {{ 0}}$ if ${\rm{E}}\{ {{{\bm f}^{\rm{T}}} { \bm g} } \} = {{0}}$. 
\end{lemma}

 We can also define LLN-orthogonality for matrices ${\bm F} = [ {{{\bm F}_{|1}}, \cdots ,{{\bm F}_{|J}}} ]  \in {{\mathbb R}^{N \times J}}$ and ${\bm G} = [ {{{\bm G}_{|1}}, \cdots ,{{\bm G}_{|J}}} ]  \in {{\mathbb R}^{N \times J}}$, denoted as $\tfrac{1}{N}{{\bm F}^{\rm{T}}}{\bm G}\stackrel{\text{LLN}}{\longrightarrow}{\bm 0}^{J \times J}$, if any pair of columns of ${\bm F}$ and ${\bm G}$, i.e., ${\bm F}_{|i}$ and ${\bm G}_{|j}$ are LLN-orthogonal to each other. 
    
    \begin{lemma} \label{Lem:LLN_MMV}
Let ${\bm G}  \!\!=\! \psi ({\bm F} )$ be PIID-separable and ${\bm F}$ is CPIID. Then $\tfrac{1}{N}{\bm F}^{\rm{T}}{{\bm G}}\!\stackrel{\text{LLN}}{\longrightarrow}\! {{\bm 0}^{J \!\times\! J}}$ provided that ${\rm{E}}\{\! {{{\bm F} ^{\rm{T}}}{\bm G}}\!\} \!=\! {{\bm 0}^{J \!\times\! J}}$.
\end{lemma}

Lemma~\ref{Lem:LLN_MMV} will be useful when we discuss port-wise orthogonality later (see Sec.~\ref{Sec:AOAMP}). 


\section{OAMP for Multiple Measurement Vector Problems} \label{Sec:MMV_system}  

\subsection{Generic Iterative Processing}\label{Sec:GIP}
In this section, we consider the problem in Fig. \ref{Fig:MMV_net}(a), where ${\bm V}$ is $N\times N$ Haar, and ${\bm \Xi}$ and ${\bm X}$ are both of size $N \times M$. The problem can be expressed concisely as: 
\BE\label{Eqn:X_Vx_MMV}
\bf{\Xi}=\bf{VX},\;\bf{\Xi}\sim \Gamma,\;\bf{X}  \sim \Phi. 
\EE 

Similar to \cite{Lei_TSP_I}, our aim is to find the {\textit{a-posteriori}} mean of $\bf{X}$ for the system in \eqref{Eqn:X_Vx_MMV}, as given below:
\BE\label{Eqn:post_mean}
\hat{\bf{X}} \equiv {\mr{E}}\{\bf{X}|\bf{\Xi}=\bf{V}\bf{X}, \Gamma, {\Phi}  \}. 
\EE
Due to complexity concerns, we focus on the following sub-optimal message passing approach.

\noindent\textbf{Generic Iterative Process (GIP) for an MMV problem:}
\BS\label{Eqn:Ite_Proc_para}\begin{align}
\!\!\!\!\mr{Initialisation}: \;&t=1,\;{\bm { { \Xi}}}^{{\rm{in}},0}= {\bm { X}}^{{\rm{in}},0} = {{\bm 0}^{N \times M}},\\
\!\!\!\!\mr{Local\;process}: \;& \bf{{{\Xi}}}^{{\mr{out}},t}\!\!=\! \gamma^t\big(\bf{{{\Xi}}}^{{\mr{in}},{t-1}}\big), {\bf{X}}^{{\mr{out}},{t}} \!\!=\! \phi^t\big({\bf{X}}^{{\mr{in}},{t-1}}\big),\\
\!\!\!\!\mr{Update}:  \;&\!\bf{{{\Xi}}}^{{\mr{in}},t} \!=\!\bf{V}{\bf{X}}^{{\mr{out}},{t}}\!\!,  \quad\; \;{\bf{X}}^{{\mr{in}},t}\!=\!\bf{V}^{\mr{T}} \bf{{{\Xi}}}^{{\mr{out}},{t}}\!\!,\label{Eqn:Ite_Proc_para2}
\end{align}\ES
Here $\gamma^t(\cdot)$ and $\phi^t(\cdot)$ are two local estimators. Their inputs and outputs, referred to as messages, are estimates of ${\bm {\Xi}}$ and ${\bm X}$ respectively. Each of $\gamma^t(\cdot)$ and $\phi^t(\cdot)$ refines its input using one of the two constraints in \eqref{Eqn:X_Vx_MMV}: $\gamma^t(\cdot)$ uses $\Gamma$ and $\phi^t(\cdot)$ uses $\Phi$. The designs of $\gamma^t(\cdot)$ and $\phi^t(\cdot)$ are usually much easier than the design of an overall estimator using $\Gamma$ and $\Phi$ jointly. 

In each iteration, the two local estimators exchange messages. Our hope is that the estimates can be refined in this way. For this purpose, we need to know the distributions of the inputs to $\gamma^t(\cdot)$ and $\phi^t(\cdot)$, which is in general a difficult task. The orthogonal principle discussed in Sec. \ref{Sec:MMV_OAMP} resolves this difficulty.



\subsection{GS Model in the MMV Case}\label{Sec:GS_model} 
Let $\hat{\bf{X}}$ be an arbitrary observation of $N\times M$ matrix $\bf{X}$. Denote
\BE\label{Eqn:GP_Gmodel_a} 
\bf{\Theta}=[\mr{E}\{\bf{X}^{\mr{T}}{\bf{X}}\}]^{-1}\mr{E}\{\bf{X}^{\mr{T}}\hat{{\bf{X}}}\},
\EE
and ${\bf{Z}} = \hat{\bf{X}} - \bf{X\Theta} $, where ${\bf{Z}}$ and $\bf{\Theta}$ are $N\times M$ and $M\times M$ matrices separately \cite{Schmidt1908,Price1958}. We write
\BE \label{Eqn:GP_Gmodel} 
\hat{\bf{X}} = \bf{X\Theta} + {\bf{Z}}.
\EE
We call \eqref{Eqn:GP_Gmodel} the Gram-Schmidt (GS) model of $\hat{\bf{X}}$ and call ${\bm Z}$ the GS error \cite{Yiyao2020}. It can be verified that ${\bf{X}}$ and ${\bf{Z}}$ are orthogonal to each other, i.e.,  
\BE\label{Eqn:GP_Gmodela}
\mr{E}\{\bf{X}^{\mr{T}}{\bf{Z}}\}\overset{\rm a.s.}{\to}\bf{0}^{M\times M}. 
\EE
Let ${\bm \Sigma }= {N^{ - 1}}{\rm{E}}\{ {{{\bm Z} ^{\rm{T}}}{\bm Z} } \}$ be the $M \times M$ covariance matrix of ${\bm Z}$. We will call ${\bm \Theta}$ and ${\bm \Sigma}$ the GS parameters of ${\hat {\bm X}}$. 

Let $\hat {\bm {\Xi}} = {\bm V}{\hat {\bm X}}$ be an estimate of ${\bm {\Xi}} = {\bm V}{{\bm X}}$. It can be verified that the GS model of ${\hat {\bm {\Xi}}}$ is $\hat {\bm {\Xi}} = {\bm {\Xi}} {\bm \Theta} + {\bm Z} '$ where ${\bm Z} ' = {\bm V}{\bm Z }$ and ${\rm{E}}\{ {{\{ {{\bm Z }'}\}^{\rm{T}}}{\bm Z }'}\} = {\rm{E}}\{ {{{\bm Z }^{\rm{T}}}{\bm Z} } \}$. Hence, we have the following.
\begin{lemma} \label{Lem:sameGS}Let ${\hat {\bm {\Xi}}}$ and ${\hat {\bm X}}$ be respectively the observations of ${\bm {\Xi}}$ and ${\bm X}$ where $\hat {\bm {\Xi}} = {\bm V}{\hat {\bm X}}$, ${\bm {\Xi}} = {\bm V}{{\bm X}}$ and ${\bm V}$ is orthogonal. Then ${\hat {\bm {\Xi}}}$ and ${\hat {\bm X}}$ have the same GS parameters, although the error terms ${\bm Z} ' $ and ${\bm Z} $ generally have different distributions. 
\end{lemma}

Lemma \ref{Lem:sameGS} will be used in Sec. \ref{Sec:MMV_SE} for the evolution analysis for OAMP-MMV.  

\subsection{OAMP-MMV Principle}\label{Sec:MMV_OAMP}
We express the messages in \eqref{Eqn:Ite_Proc_para} in their GS models as
\begin{subequations} \label{Eqn:Ite_Proc_GS}
\begin{align}
{\bm {\Xi}}^{{\rm{out}},t} &\!=\! {\bm {\Xi}}{\bm \Theta }_{{\Xi}}^{{\rm{out}},t}\! + \!{\bm Z }_{{ \Xi}}^{{\rm{out}},t}, {\bm X}^{{\rm{out}},t}\! =\! {\bm X}{\bm \Theta} _{{ X}}^{{\rm{out}},t} \!+\! {\bm Z }_{{ X}}^{{\rm{out}},t},\\
 {\bm {\Xi}}^{{\rm{in}},t} &= {\bm {\Xi}}{\bm \Theta }_{{\Xi}}^{{\rm{in}},t} + {\bm Z }_{{ \Xi}}^{{\rm{in}},t},\; {\bm X}^{{\rm{in}},t} = {\bm X}{\bm \Theta} _{{ X}}^{{\rm{in}},t} + {\bm Z }_{{ X}}^{{\rm{in}},t}.
\end{align}
\end{subequations}
Let ${\bm \Sigma }_{{ \Xi}}^{{\rm{out}},t}$,${\bm \Sigma }_{{ \Xi}}^{{\rm{in}},t}$,${\bm \Sigma }_{{ X}}^{{\rm{out}},t}$ and ${\bm \Sigma }_{{ X}}^{{\rm{in}},t}$ be the $M \times M$ auto-covariance matrices of the GS errors ${\bm Z }_{{ \Xi}}^{{\rm{out}},t},{\bm Z }_{{ \Xi}}^{{\rm{in}},t},{\bm Z }_{{ X}}^{{\rm{out}},t}$ and ${\bm Z }_{{ X}}^{{\rm{in}},t}$ respectively. Then, according to Lemma \ref{Lem:sameGS}, the following relationships hold when ${\bm V}$ is orthogonal:
\begin{subequations} \label{Eqn:Ite_Proc_GSa}
\begin{align}
{\bm \Theta} ^t_{\gamma}  ={\bm \Theta }_{{\Xi}}^{{\rm{out}},t} = {\bm \Theta} _{{X}}^{{\rm{in}},t},\;\;{\bm \Theta }^t_{\phi}  = {\bm \Theta} _{{X}}^{{\rm{out}},t} = {\bm \Theta} _{{{ \Xi}}}^{{\rm{in}},t},\\
{\bm \Sigma}^t _{\gamma}  ={\bm \Sigma }_{{\Xi}}^{{\rm{out}},t} = {\bm \Sigma} _{{X}}^{{\rm{in}},t},\;\;{\bm \Sigma }^t_{\phi}  = {\bm \Sigma} _{{X}}^{{\rm{out}},t} = {\bm \Sigma} _{{{ \Xi}}}^{{\rm{in}},t}.
\end{align}
\end{subequations}
We will use \eqref{Eqn:Ite_Proc_GSa} to simplify the derivations of SE in Sec.~\ref{Sec:MMV_SE}. 

Let ${\bm Z }_{{ \Xi}}^{{\rm{out}},0} = {\bm Z }_{{X}}^{{\rm{out}},0}= {{\bm 0}^{N \times M}}$. The matrices defined below consist of all the GS errors in \eqref{Eqn:Ite_Proc_GS}, i.e., ${\bm Z}_\Xi ^{{\rm{in}},t'}$, ${\bm Z}_\Xi ^{{\rm{out}},t'}$, ${\bm Z}_X^{{\rm{in}},t'}$ and ${\bm Z}_X^{{\rm{out}},t'}$, for $t'=0,1,\cdots,t$:
\BS\label{Eqn:Error_matrix}\begin{alignat}{2}
\!\!\!\!\!\!{\bm Z}_\Xi ^{{\rm{in}},\{ t \}} &\!\!\equiv\! [\! {{\bm Z}_\Xi ^{{\rm{in}},0}, \!\cdots\! ,{\bm Z}_\Xi ^{{\rm{in}},t }} ],
{\bm Z}_{\Xi}^{{\rm{out}},\{ t\}} &\!\equiv\! [\! {{\bm Z}_{\Xi}^{{\rm{out}},0}, \!\cdots\! ,{\bm Z}_{\Xi}^{{\rm{out}},t }} ] ,\\
\!\!\!\!\!\!{\bm Z}_X ^{{\rm{in}},\{ t \}} &\!\equiv\! [\! {{\bm Z}_X ^{{\rm{in}},0}, \!\cdots\! ,{\bm Z}_X ^{{\rm{in}},t }} ],
{\bm Z}_X^{{\rm{out}},\{ t \}} &\!\!\equiv \![\! {{\bm Z}_X^{{\rm{out}},0}, \!\cdots\! ,{\bm Z}_X^{{\rm{out}},t }} ].\label{Eqn:Error_matrixc}
\end{alignat}\ES
From \eqref{Eqn:X_Vx_MMV}, \eqref{Eqn:Ite_Proc_para} and \eqref{Eqn:Error_matrix}, we have the following constraint:
\BE\label{Eqn:G=VF}
\!\!\![ {{\bm \Xi },\!{\bm Z}_\Xi ^{{\rm{out}},\{ t-1 \}},\!{\bm Z}_\Xi ^{{\rm{in}},\{ t-1 \}}} ] \!=\! {\bm V}[ {{\bm X},\!{\bm Z}_X^{{\rm{in}},\{ t-1 \}},\!{\bm Z}_X^{{\rm{out}},\{ t-1 \}}} ].
\EE
\begin{definition}[OAMP-MMV] \label{Def:OAMP-MMV}
 OAMP-MMV is a special case of GIP-MMV in \eqref{Eqn:Ite_Proc_para} when the following LLN-orthogonality holds for $N \to \infty $, $t \ge 1$ and $0 \le t' < t$,
\begin{subequations} \label{Eqn:Orthogonality}
 \begin{align}
\!\!\!\!\!\!\!{\tfrac{1}{N}( {\bm Z}_\Xi ^{{\rm{in}},t'} )^{\rm{T}}}{\bm Z}_\Xi ^{{\rm{out}},t} &\stackrel{\text{LLN}}{\longrightarrow} {\bm 0}^{M \times M},{\tfrac{1}{N}( {\bm Z}_X ^{{\rm{in}},t'} )^{\rm{T}}}{\bm Z}_X ^{{\rm{out}},t} \stackrel{\text{LLN}}{\longrightarrow} {\bm 0}^{M\times M},\\
\!\!\!\!\!\!\!\!\!\!\!{\tfrac{1}{N}{\bm {\Xi}}^{\rm{T}}}{\bm Z}_\Xi ^{{\rm{out}},t} &\!\stackrel{\text{LLN}}{\longrightarrow}\! {\bm 0}^{M \times M},{\tfrac{1}{N}{\bm X}^{\rm{T}}}{\bm Z}_X ^{{\rm{out}},t} \!\stackrel{\text{LLN}}{\longrightarrow}\! {\bm 0}^{M \times M}.
\end{align} 
 \end{subequations}
\end{definition}
We will discuss the designs of $\gamma^t(\cdot)$ and $\phi^t(\cdot)$ to meet the above constraints in Sec. \ref{Sec:MMV_OAMP}.

\subsection{Error Behaviour of OAMP-MMV} \label{Sec:Error_OAMPMMV}
We first introduce some notations. Return to \eqref{Eqn:G=VF}. Define
\begin{subequations} \label{Eqn:Error_trans}
\begin{align}
{{\bm A}_*^t} &= \big[{\bm \Xi} ,{\bm Z}_\Xi ^{{\rm{out}},\{ t-1 \}}\big],\;\;\;&&{{\bm B}_*^t} = \big[{\bm X},{\bm Z}_X^{{\rm{in}},\{ t-1 \}}\big], \label{Eqn:Error_trans1}\\
{{\bm A}_{**}^t } &= {\bm Z}_\Xi ^{{\rm{in}},\{ t-1 \}},\;\;\;&&{{\bm B}_{**}^t } = {\bm Z}_X ^{{\rm{out}},\{ t-1 \}},\label{Eqn:Error_trans2}\\
 {\bf A}^t &= [ {{{\bf A}_ *^t },\;\;{{\bf A}_{ *  * }^t}} ],  &&{\bf B}^t = [ {{{\bf B}_ *^t },\;\;{{\bf B}_{ *  * }^t}} ].\label{Eqn:Error_trans3}
\end{align}
\end{subequations}
From \eqref{Eqn:G=VF} and \eqref{Eqn:Error_trans}, we have ${{\bm A}_*^t } = {\bm V}{{\bm B}_*^t }$ and ${{\bm A}_{**}^t } = {\bm V}{{\bm B}_{**}^t }$. Then 
\begin{equation} \label{Eqn:Error_A=VB}
{\bf A}^t = {\bf V}{\bf B}^t.
\end{equation}
We now analyze OAMP-MMV based on the following assumptions.

\begin{assumption} \label{Asu:MMV}
(i) At the beginning of each experiment, $\bf V$ is independently sampled using ${\bf V} \sim {\bm{\mathcal H}^{N \times N}}$. (ii) During each experiment, $\bf V$ remains unchanged for all iterations. (iii) Both $\gamma^t(\cdot)$ and $\phi^t(\cdot)$ are PIID-separable and independent of $\bm V$. 
\end{assumption}

Equivalent forms of Assumption \ref{Asu:MMV} have been used in almost all AMP-type algorithms (including OAMP). Note that the assumption of Haar matrices is less restrictive of IIDG sensing matrices for AMP \cite{Donoho2009,Bayati2011}. In fact, an IIDG matrix is a special case of a Haar matrix when the $k^{\rm{th}}$ moment ${M^{ - 1}}{\rm{Tr}}\{ {{{( {D{D^{\rm{T}}}})^k}}}\}$ converges almostly surely to the Marchenko-Pastur Distribution in the large system limit \cite{Tulino2004}. 

From \eqref{Eqn:Error_A=VB}, Assumption \ref{Asu:MMV} and the Bolthausen’s conditioning technique \cite{Bayati2011,Takeuchi2020,Bolthausen2014}, we can see that, at the beginning of iteration $t$ of each experiment, $\bf V$ is uniformly distributed over the subset of the sample space of ${\bf V} \sim {{\bm{\mathcal H}}^{N \times N}}$ constrained by ${\bf A}^t={\bf V}{\bf B}^t$ in \eqref{Eqn:Error_A=VB}. This constraint means that ${\bf V}$ should be consistent with all the errors up to iteration $t-1$. We denote the distribution of such ${\bf V}$ as
\begin{equation} \label{Eqn:Vconstrained_MMV}
    {\bf V} \sim {\bm {\mathcal H}}( {\bf A}^t = {\bf V}{\bf B}^t ).
\end{equation}
For such $\bf V$, Lemma \ref{Lem:generalV} is not directly applicable. For example, we cannot apply Lemma \ref{Lem:generalV} to obtain the distribution of ${\bm Z}_\Xi ^{{\rm{in}},t} = {\bm V}{\bm Z}_X^{{\rm{out}},t}$. (For more details, please see Appendix A.) 

We say that a matrix $\bm F$ is CPIIDG and row-wise joint-Gaussian (CPIIDG-RJG) if its every column is PIIDG and every row is joint Gaussian. Clearly, Lemma~\ref{Lem:CPIID_1} is applicable to a CPIIDG-RJG matrix. We also define the angles of $\bm F$ as $\{{\bm f}/\|\bm f\|, \forall {\bm f} \in {\rm{Clmn}}(\bm F)\}$. 

\begin{theorem} \label{Theo:IIDG_MMV}
Under Assumption \ref{Asu:MMV}, let OAMP-MMV be initialized with ${\bm {{\Xi} }}^{{\rm{in}},0}= {\bm { X}}^{{\rm{in}},0} = {{\bm 0}^{N \times M}}$. Then the following claims hold for the errors in \eqref{Eqn:Error_matrix} at any finite $t$ where $N \to \infty $:
\begin{itemize}
\item[(i)] both ${\bm Z}_\Xi ^{{\rm{in}},\{ t -1\}}$ and ${\bm Z}_X ^{{\rm{in}},\{ t-1 \}}$ converge to CPIIDG-RJG in distribution;
\item[(ii)] the angles of ${\bm Z}_\Xi ^{{\rm{in}},\{ t -1\}}$ and ${\bm Z}_X ^{{\rm{in}},\{ t-1 \}}$ are asymptotically entry-wise independent of $\bm z$ in the Markov chain below:
\BE \label{Eqn:Markov_MMV}
\!\!\!\!{\bm z} \!\to\! [{\bm Z}_{\Xi}^{{\rm{out}},t},{\bm Z}_X^{{\rm{out}},t},{{\bm A}^t,{\bm B}^t}] \!\to\! {\bm V} \!\sim\! {\bm {\mathcal H}}({ {{\bm A}^t \!=\! {\bm V}{\bm B}^t} }). 
\EE 


\end{itemize}
 \end{theorem}
\begin{IEEEproof}
We will first explain the conditions. In Assumption~\ref{Asu:MMV}, (i) ensures that ${\bm V}$ is Haar; (ii) ensures that ${\bm V}$ is constrained by \eqref{Eqn:Error_A=VB}; and (iii) ensures that there is no additional constraint on ${\bm V}$ other than \eqref{Eqn:Error_A=VB}. (Note: Additional constraints may result $\gamma^t(\cdot)$ and $\phi^t(\cdot)$ involving ${\bm V}$.)

From \eqref{Eqn:Ite_Proc_para} and the initialization, we have ${\bm { { \Xi}}}^{{\rm{in}},0}= {\bm { X}}^{{\rm{in}},0} = {{\bm 0}^{N \times M}}$. Their GS models are respectively ${{\bm \Xi} ^{{\rm{in}},0}} = {\bm \Xi}  \cdot {{\bm 0}^{M \times M}} + {{\bm 0}^{N \times M}}$ and ${{\bm X} ^{{\rm{in}},0}} = {\bm X}  \cdot {{\bm 0}^{M \times M}} + {{\bm 0}^{N \times M}}$, in which the GS errors (both ${\bm 0}^{N\times M}$) are special cases of CPIIDG-RJG and the angles are entry-wise independent of arbitrary $\bm z$ for $\bm z$ in \eqref{Eqn:Markov_MMV}. Hence (i) and (ii) hold at $t=1$.

We now prove Theorem \ref{Theo:IIDG_MMV} by induction. Assume that Theorem \ref{Theo:IIDG_MMV} holds for any $t>0$ and consider $t+1$. Consider applying Lemma~\ref{Chap2:Lem:P=VQ_perp} in Appendix A to ${\bm A}^t={\bm V}{\bm B}^t$ in \eqref{Eqn:Error_A=VB}. We first verify Assumption~\ref{Asu:generalHaar} in Appendix A. 
\begin{itemize}
\item[(a)] From the induction assumption, ${{\bm A}_{ *  * }} ={\bm Z}_\Xi^{{\rm{in}},\{ t-1 \}}$ converges to CPIIDG-RJG. 
\item[(b)] From \eqref{Eqn:Orthogonality}, we have $\tfrac{1}{N}{[ {{\bm X},{\bm Z}_X^{{\rm{in}},\{ t-1 \}}} ]^{\rm{T}}}{\bm Z}_X^{{\rm{out}},t} \!\stackrel{\text{LLN}}{\longrightarrow} \!{\bm 0}$ in probability. Combining this with \eqref{Eqn:Error_trans} and from Lemma~\ref{Lem:LLN_MMV}, we have
\BE
 \!\!\!\!\! \!\!\!  {\rm E}\{\!{( {{{\bm B}_ *^t }} )^{\rm{T}}}{\bm Z}_X^{{\rm{out}},t}\} \!\!=\! {\bm 0}^{(t+1)M \!\times\! M} \!\!\Rightarrow \!\! {\rm{Sp}}(\!{\bm Z}_X^{{\rm{out}},t}  )\!\!\perp\!{\rm{Sp}}( \!{{{\bm B}_*}} ).
\EE
\item[(c)] Since $\gamma^{t'}(\cdot)$, $\forall t'\!\le\! t$ is PIID-separable and ${\bm \Xi}$ is CPIID, ${\bm Z}_{\Xi}^{{\rm{out}},t'}\!\!=\!\!\gamma^{t'}({\bm Z}_{\Xi}^{{\rm{in}},t'-1})\!\!-\!{\bm \Xi}{\bm \Theta}_{\Xi}^{{\rm{out}},t'}$ is CPIID and thus ${\bm A}^t\!=\! [{\bm \Xi}, {\bm Z}_{\Xi}^{{\rm{out}},\{t-1\}},{\bm Z}_{\Xi}^{{\rm{in}},\{t-1\}}]$ in \eqref{Eqn:Error_trans} is also CPIID. 
\end{itemize}

The above shows that Assumption~\ref{Asu:generalHaar} holds for ${\bm A}^t\!=\!{\bm V}\!{\bm B}^t$. Then from Lemma~\ref{Chap2:Lem:P=VQ_perp}(i) in Appendix~\ref{APP:ConstrainedHaar}, ${\bm Z}_\Xi ^{{\rm{in}},\{ {t } \}} \!\!=\!\! [ {{{\bm A}_{**}^t},{\bm V}{\bm Z} _X^{{\rm{out}},t}} ]$ converges to CPIIDG-RJG. Following Lemma~\ref{Chap2:Lem:P=VQ_perp}(ii), the angles of ${\bm Z}_\Xi ^{{\rm{in}},\{ {t } \}}$ are asymptotically entry-wise independent of $\bm z$. Due to the symmetry of the problem, we can prove the two claims for ${\bm Z}_X ^{{\rm{in}},\{ {t } \}}$ in a similar way. Hence the induction holds for $t+1$. This proves Theorem \ref{Theo:IIDG_MMV}.
\end{IEEEproof}

 \subsection{State Evolution}   \label{Sec:MMV_SE}
 The implications of Theorem \ref{Theo:IIDG_MMV} are as follows.
 
 From \eqref{Eqn:GP_Gmodel}-\eqref{Eqn:Ite_Proc_GS}, the GS parameters at the output of $\phi^t(\cdot)$ can be calculated as
 \begin{subequations} \label{Eqn:MSE_IIDG_MMV}
 \begin{align}
\!\!\!\!\bf{\Theta}_{\phi}^{t} &\!=\! {[ {\mathop {\rm{E}}\limits_{\bm X} \{ {{{\bm X}^{\rm{T}}}{\bm X}} \}} ]^{ - 1}}\!\mathop {\rm{E}}\limits_{\bm X,{\bm V}} \{ {{{\bm X}^{\rm{T}}}{\phi ^t}( {{\bm X}\bf{\Theta}_{\gamma}^{t-1}+ {\bm Z}_X^{{\rm{in}},t - 1}} )} \},\\
\!\!\!\! \bf{\Sigma}_{\phi}^{t} &\!=\! \mathop {{\rm{ACF}}}\limits_{\bm X,\bm V} \{ {{\phi ^t}( {{\bm X}\bf{\Theta}_{\gamma}^{t-1} + {\bm Z}_X^{{\rm{in}},t - 1}} ) - {\bm X}{\bm \Theta} _{\phi}^{t}} \},
 \end{align}
 \end{subequations}
 where the average is over the joint sample space of ${\bm X}$ and ${\bm V} \sim {\bm{\mathcal H}}( {{{\bm A}^t} = {\bm V}{{\bm B}^t}} )$.
 
 From Theorem \ref{Theo:IIDG_MMV}, ${{\bm Z}_X^{{\rm{in}},t - 1}}$ is asymptotically CPIIDG-RJG and entry-wise independent of arbitrary ${\bm z}$ for $\bm z$ in \eqref{Eqn:Markov_MMV}. Examples of such ${\bm z}$ include ${\bm \Xi}$, ${\bm X}$ and additive thermal noise in the system (i.e., ${\bm A}^1={\bm \Xi}$ and ${\bm B}^1={\bm X}$).
Therefore,  we can rewrite \eqref{Eqn:MSE_IIDG_MMV} as 
 \begin{subequations} \label{Eqn:MSE_phi_IDDG_MMV}
 \begin{align}
 \!\!\!\!\!\!\!\!\!\bf{\Theta}_{\phi}^{t}&\!\!=\!\! \big[\mathop\mr{E}\limits_{ 
\bf{X}}\{\bf{X}^{\mr{T}}{\bf{X}}\}\big]^{-1} \!\! \mathop\mr{E}\limits_{\bf{X},\bf{\mathcal Z}^{t-1}_\gamma} \!\! \big\{ \bf{X}^{\mr{T}} \phi^t(\bf{X}\bf{\Theta}_{\gamma}^{t-1} \!\!+\! \!{\bf{\mathcal Z}}^{t-1}_\gamma)\big\}, \\
\!\!\!\!\! \bf{\Sigma}_{\phi}^{t}&  \!\! =\!  \mathop\mr{ACF}\limits_{\bm X,\bf{\mathcal Z}^{t-1}_\gamma}\! \big\{ \phi^t(\bf{X}\bf{\Theta}_{\gamma}^{t-1} \!+\!\bf{\mathcal Z}^{t-1}_\gamma)\!-\! \bf{X}\bf{\Theta}_{\phi}^{t} \big\}\!,  
\end{align}
\end{subequations}
where each row of $\bf{\mathcal Z}^{t-1}_\gamma$ is IID drawn from distribution $\mathcal{N}(\bf{0},\bf{\Sigma}_{\gamma}^{t-1})$. Similarly, the GS parameters of $\gamma^t(\cdot)$ can be calculated as
\begin{subequations}  \label{Eqn:MSE_gamma_IDDG_MMV}
\begin{align}
 \!\!\!\!\bf{\Theta}_{\gamma}^{t} 
&\!\!=\! \big[\mathop\mr{E}\limits_{ 
\bf{\Xi}}\{\bf{\Xi}^{\mr{T}}{\bf{\Xi}}\}\big]^{-1} \!\! \! \mathop\mr{E}\limits_{\bf{\Xi}, \bf{\mathcal{Z}}^{t-1}_\phi} \!  \big\{ \bf{\Xi}^{\mr{T}} \gamma^t(\bf{\Xi}\bf{\Theta}_{\phi}^{t-1} \!+\! \!{\bf{\mathcal{Z}}}^{t-1}_\phi)\big\},\\
\!\!\bf{\Sigma}_{\gamma}^{t} \!& \!=\!   \mathop\mr{ACF}\limits_{ \bf{\Xi},\bf{\mathcal{Z}}^{t-1}_\phi}\!\big\{ \gamma^t(\bf{\Xi}\bf{\Theta}_{\!\phi}^{t-1} \!+\!\bf{\mathcal{Z}}^{\!t-1}_\phi)\!-\! \bf{\Xi}\bf{\Theta}_{\gamma}^{t} \big\}\!,
 \end{align}
 \end{subequations}
with each row of $\bf{\mathcal{Z}}^{t-1}_\phi$ IID drawn from $\mathcal{N}(\bf{0},\bf{\Sigma}_{\phi}^{t-1})$. 

The key in \eqref{Eqn:MSE_phi_IDDG_MMV} and \eqref{Eqn:MSE_gamma_IDDG_MMV} is to replace ${\bm Z}_X^{{\rm{in}},t-1}$ and ${\bm Z}_\Xi^{{\rm{in}},t-1}$ by additive Gaussian noise samples ${\bm {\mathcal Z}}_\gamma^{t-1}$ and ${\bm {\mathcal Z}}_\phi^{t-1}$ respectively, which makes the problem trackable. As the result, the input and output GS parameters can be expressed as a recursion in \eqref{Eqn:SE_MMV} below, where the functions 
$\gamma^t_{\mr{SE}}(\cdot)$ and $\phi^t_{\mr{SE}}(\cdot)$ 
may or may not have explicit expressions, but they can always be numerically evaluated using \eqref{Eqn:MSE_phi_IDDG_MMV} and \eqref{Eqn:MSE_gamma_IDDG_MMV}.

\underline{State evolution for OAMP-MMV:} Starting with $t=1$ and $\bf{\Theta}^{0}_\gamma=\bf{\Theta}^{0}_\phi={\bm \Sigma}_{\gamma}^0={\bm \Sigma}_{\phi}^0=\bf{0}^{M \times M}$,  
\BS \label{Eqn:SE_MMV} \begin{align}
\big(\bf{\Theta}^{t}_\gamma, \bf{\Sigma}_{\gamma}^{t}\big)&= \gamma_{\mr{SE}}^t\big(\bf{\Theta}^{t-1}_\phi, \bf{\Sigma}_{\phi}^{t-1}\big),\\
\big(\bf{\Theta}^{t}_\phi, \bf{\Sigma}_{\phi}^{t}\big) &=\phi_{\mr{SE}}^t\big(\bf{\Theta}^{t-1}_\gamma, \bf{\Sigma}_{\gamma}^{t-1}\big).
\end{align}\ES
 
\subsection{Gram-Schmidt Orthogonalization (GSO) for MMV} \label{Sec:MMV_GSO}
We now discuss the realization of \eqref{Eqn:Orthogonality} required for OAMP-MMV. Start from an arbitrary $\hat \psi(\cdot)$. We outline a GSO procedure to construct $\psi(\cdot)$ with input-output error orthogonality.
\begin{definition}
 Let ${\bm X}^{\rm{out}}$ and ${\bm X}^{\rm{in}}$ be two observations of ${\bm X}\! \in\! {\mathbb{R}^{N \times M}}$ with their GS models given by: ${\bm X}^{\rm{out}} \! =\! {\bm X}{{\bm \Theta} ^{{\rm{out}}}}\! + \!{{\bm Z} ^{{\rm{out}}}}$ and ${\bm X}^{\rm{in}} \!=\! {\bm X}{{\bm \Theta} ^{{\rm{in}}}} \!+\! {{\bm Z} ^{{\rm{in}}}}$. We say that ${\bm X}^{\rm{out}}  \!=\! { \psi }( {\bm X}^{\rm{in}} )$ is an orthogonal estimator if ${\rm{E}}\{ {{( {{{\bm Z }^{{\rm{in}}}}} )^{\rm{T}}}{{\bm Z }^{{\rm{out}}}}} \} \!=\! {{\bm 0}^{M \times M}}$.
\end{definition} 

Let $\hat{\bm X}^{\rm{out}}    = \hat \psi ( {\bm X}^{\rm{in}}   )$ be an arbitrary prototype estimator. We can construct an orthogonal $ \psi ( {\bm X}^{\rm{in}}  )$ as follows:
\BE\label{Eqn:GGSO}
{\bm X}^{\rm{out}}  =\psi({\bm X}^{\rm{in}}  )=\hat{\psi}({\bm X}^{\rm{in}}  )-  {\bm X}^{\rm{in}}  \bf{\Delta}.  
\EE
The orthogonal requirement can be rewritten as:
\begin{subequations}  \label{Eqn:GGSO_orth}
\begin{align}
\!\!\mr{E}\{(\bf{Z}^{\mr{in}})^{\mr{T}}\bf{Z}^{\mr{out}}\} &\!=\!\mr{E}\{(\bf{Z}^{\mr{in}})^{\mr{T}}({\bm X}^{\rm{out}} -\bf{X\Theta}^{\mr{out}})\} \\
&\! =\!\mr{E}\{(\bf{Z}^{\mr{in}})^{\mr{T}}{\bm X}^{\rm{out}}  \} \\
&\!=\!\mr{E} \{(\bf{Z}^{\mr{in}})^{\mr{T}}\big(\hat{\psi}({\bm X}^{\rm{in}}) \!-\! {\bm X}^{\rm{in}}\bf{\Delta}\big)  \}={\bm 0}. \label{Eqn:GGSO_orthd}
\end{align}
\end{subequations}
Notice that
\BE \label{Eqn:GGSO_equ}
\!\!\! \!\! \mr{E}\{\!(\!\bf{Z}^{\mr{in}})^{\!\mr{T}}{\bm X}^{\rm{in}}  \!\} \!=\!\mr{E}\{\!(\!\bf{Z}^{\mr{in}})^{\!\mr{T}}{(\bf{X\Theta}^{\mr{in}}\!\!+\!\bf{Z}^{\mr{in}})}\!\} \!=\!\mr{E}\{\!(\!\bf{Z}^{\mr{in}})^{\!\mr{T}}\bf{Z}^{\mr{in}}\}.
\EE
Combining \eqref{Eqn:GGSO_orthd} and \eqref{Eqn:GGSO_equ}, we have
\BE\label{Eqn:GGSO_B}
\bf{\Delta} = [\mr{E}\{(\bf{Z}^{\mr{in}})^{\mr{T}}\bf{Z}^{\mr{in}}\}]^{-1}\mr{E}\{ (\bf{Z}^{\mr{in}})^{\mr{T}}\hat{\psi}({\bm X}^{\rm{in}}  )\}. 
\EE
\subsection{OAMP-MMV via GSO}
Using \eqref{Eqn:GGSO}-\eqref{Eqn:GGSO_B}, we can construct local estimators with un-correlated input and output errors, which is weaker than the LLN-orthogonal requirement in \eqref{Eqn:Orthogonality}. Theorem \ref{Theo:GSO_OAMP_MMV} below addresses this issue. 
\begin{theorem}\label{Theo:GSO_OAMP_MMV}
The LLN-orthogonality in \eqref{Eqn:Orthogonality} holds if $\gamma^t(\cdot)$ and $\phi^t(\cdot)$, $\forall t$ are orthogonal and PIID-separable.
\end{theorem}
\begin{IEEEproof}  
We prove by induction on $t$. Using Lemma~\ref{Lem:LLN_MMV}, we can show that  \eqref{Eqn:Orthogonality} holds for $t=1$ under Assumption \ref{Asu:MMV}. Now we assume that \eqref{Eqn:Orthogonality} holds for $t$ ($t>0$) and consider $t+1$. The induction assumption implies that Theorem \ref{Theo:IIDG_MMV} holds at $t+1$ (since \eqref{Eqn:Orthogonality} is the condition for Theorem \ref{Theo:IIDG_MMV}). Then ${\bm Z}_X^{{\rm{in}},t}$ and ${\bm Z}_X^{{\rm{in}},t'}$, $t'=0,1,\!\cdots\!,t-1$ are CPIIDG-RJG. Denote 
\begin{equation}
{\bm Z}  = {\bm Z}_X^{{\rm{in}},t }{\bm \Theta } - {\bm Z}_X^{{\rm{in}},t'}, \label{Eqn:Error_define}
\end{equation}
where 
\begin{equation} 
  {\bm \Theta}  = {[ {{\rm{E}}\{ {{( {\bm Z}_X^{{\rm{in}},t } )^{\rm{T}}}{\bm Z}_X^{{\rm{in}},t }} \}} ]^{ - 1}}{\rm{E}}\{ {{( {\bm Z}_X^{{\rm{in}},t } )^{\rm{T}}}{\bm Z}_X^{{\rm{in}},t'}} \}.  
\end{equation} 
It can be verified that ${\rm{E}}\big\{ {{( {\bm Z}_X^{{\rm{in}},t } )^{\rm{T}}}{\bm Z} } \big\} = {\bm 0}^{M \times M}$. (This can be compared to \eqref{Eqn:GP_Gmodel_a} for the GS model \eqref{Eqn:GP_Gmodel}.)  Then ${\bm Z}$ and ${\bm Z}_X^{{\rm{in}},t }$ are uncorrelated and so mutually independent due to their Gaussianity. (As ${\bm Z}_X^{{\rm{in}},t }$ and ${\bm Z}_X^{{\rm{in}},t'}$ are CPIIDG-RJG, ${\bm Z} $ is also CPIIDG-RJG.) Due to the Markov chain ${\bm Z}  \to {\bm Z}_X^{{\rm{in}},t } \to {\bm Z}_X^{{\rm{out}},t+1}$, ${\bm Z}$ is also independent of ${\bm Z}_X^{{\rm{out}},t+1}$. Besides, ${\bm Z}_X^{{\rm{in}},t }$ and ${\bm Z}_X^{{\rm{out}},t+1 }$ are orthogonal since $\phi^{t+1}(\cdot)$ is orthogonal to each other and PIID-separable. Then we can get the following:
\begin{equation} \label{Eqn:xi_fin_normal}
{\rm{E}}\{ {{{\bm Z} ^{\rm{T}}}{\bm Z}_X^{{\rm{out}},t+1}} \} = {{\bm 0}^{M \times M}},\;{\rm{E}}\{ {{( {\bm Z}_X^{{\rm{in}},t})^{\rm{T}}}{\bm Z}_X^{{\rm{out}},t+1}} \} = {{\bm 0}^{M \times M}}.
\end{equation}
Since $\bf Z$, $ {\bm Z}_X^{{\rm{in}},t}$ are CPIID, from Lemma~\ref{Lem:LLN_MMV}, we have
\begin{equation} \label{Eqn:xi_fin_LLN}
\tfrac{1}{N}{{\bm Z} ^{\rm{T}}} {\bm Z}_X^{{\rm{out}},t+1} \stackrel{\text{LLN}}{\longrightarrow} {\bm 0}^{M \times M}, \; \tfrac{1}{N}{(  {\bm Z}_X^{{\rm{in}},t} )^{\rm{T}}} {\bm Z}_X^{{\rm{out}},t+1} \stackrel{\text{LLN}}{\longrightarrow}  {\bm 0}^{M \times M}.
\end{equation}
Then from \eqref{Eqn:Error_define} and \eqref{Eqn:xi_fin_LLN}, we can get
\begin{equation}
\tfrac{1}{N}{(  {\bm Z}_X^{{\rm{in}},t'} )^{\rm{T}}} {\bm Z}_X^{{\rm{out}},t+1} = \tfrac{1}{N}{( { {\bm Z}_X^{{\rm{in}},t}{\bm \Theta}  - {\bm Z} } )^{\rm{T}}} {\bm Z}_X^{{\rm{out}},t+1} \stackrel{\text{LLN}}{\longrightarrow} {\bm 0}^{M \times M}.
\end{equation}
with $t'=0,1,\cdots,t-1$. Similarly, with an orthogonal and PIID-separable $\gamma^{t+1}(\cdot)$, we can prove $\frac{1}{N}{({\bm Z}_\Xi ^{{\rm{in}},t'} )^{\rm{T}}}{\bm Z}_\Xi ^{{\rm{out}},t+1} \stackrel{\text{LLN}}{\longrightarrow} {\bm 0}^{M \times M}$ for any $0 \le t' \le t$. Therefore  \eqref{Eqn:Orthogonality} holds at iteration $t+1$, which completes the proof of Theorem \ref{Theo:GSO_OAMP_MMV}.
\end{IEEEproof}

The discussions in Secs. III-F and III-G provide a construction technique for OAMP-MMV in Definition~\ref{Def:OAMP-MMV}.

 \section{OAMP for Multiple Transform Problems} \label{Sec:MC}
The following discussions do not involve row indexes. Therefore, for simplicity and different from the notation in Sec. II, we drop ``$-$'', ``$|$" and ``$||$" in the subscripts for the indexes of rows, columns or subsets of columns. All subscripts are column-block related.
\subsection{Multiple Transform (MT) Problems} \label{Sec:system_model}
Fig. \ref{Fig:MMV_net}(a) involves one transform ${\bm V}$. We now consider an MT system involving multiple transforms (MT). Let $\{ {{{\bm {\Xi}}_k}} \}$ and $\{ {{{\bm X}_k}} \}$ be two sets of variables connected by Haar transforms:
\begin{equation}  \label{Eqn:system_MC}
    {{\bm {\Xi}}_k} = {{\bm V}_k}{{\bm X}_k},\;\;\;k = 1,2, \cdots ,K,
\end{equation}
where ${{\bm V}_k} \sim {\bm{\mathcal H}^{N \times N}}$, $k=1,2,\cdots, K$ are independent with each other. In general, $\{ {{{\bm {\Xi}}_k}} \}$ and $\{ {{{\bm X}_k}} \}$ can be vectors or matrices, and the matrices sizes can be different for different $k$. For simplicity, in this paper, we assume that all the elements in $\{ {{{\bm {\Xi}}_k}} \}$ and $\{ {{{\bm X}_k}} \}$ are of size $N\times M$, which can be simply defined as an MMV-MT model. 
\subsection{Multi-Port Constraints}
An MT system involves at least one multiple-port constraint defined below. Denote by ${\mathcal I}_k$ a set of integers that include $k$. For any $k'\in {\mathcal I}_k$, we write ${\mathcal I}_k={\mathcal I}_{k'}$. As examples,
\BE  \label{Eqn:Ik_example1}
{{\mathcal I}_1} = {{\mathcal I}_2} = \{ {1,2} \} \text{ and } {{\mathcal I}_3} = {{\mathcal I}_4} = {{\mathcal I}_5} = \{ {3,4,5} \}.
\EE
Denote by ${{\bm X}_{{{\mathcal I}_k}}}$ the set of all ${\bm X}_k$ with indexes in ${\mathcal I}_k$, i.e., ${{\bm X}_{{{\mathcal I}_k}}} = \{ {{{\bm X}_{k'}},k' \in {{\mathcal I}_k}} \}$. For example,
\BE   \label{Eqn:Ik_example2}
{{\bm X}_{{{\mathcal I}_1}}}={{\bm X}_{{{\mathcal I}_2}}}=\{{\bm X}_1, {\bm X}_2\} \text{ for } {{\mathcal I}_1} = {{\mathcal I}_2} = \{ {1,2} \}.
\EE  
We denote ${\Phi _{{{\mathcal I}_k}}}$ as a constraint on the elements in ${\mathcal I}_{k}$. We call ${\Phi _{{{\mathcal I}_k}}}$ a $L_k$-port constraint where $L_k=| {{{\mathcal I}_k}}|$. We also call it a multi-port and single-port constraint for $L_k>1$ and $L_k=1$ respectively. For example,
\BE   \label{Eqn:Ik_example3}
{\Phi_{{{\mathcal I}_1}}}={\Phi_{{{\mathcal I}_2}}}=\Phi_{1,2} \text{ for } {{\mathcal I}_1} = {{\mathcal I}_2} = \{ {1,2} \},
\EE
is a multi-port (more precisely, 2-port) constraint. Similarly, let ${\mathcal J}_k$ be a set of integers that include $k$. We define ${{\bm \Xi}_{{{\mathcal J}_k}}}$ and ${\Gamma_{{{\mathcal J}_k}}}$ in a similar way as ${{\bm X}_{{{\mathcal I}_k}}}$ and ${\Phi_{{{\mathcal I}_k}}}$. 

\begin{figure} 
 \centering
 \subfigure[A 2-transform system]{
\begin{minipage}[b]{0.35\textwidth} 
\includegraphics[width=\textwidth]{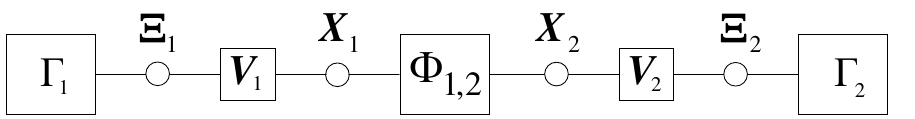} 
\end{minipage}   
} \vspace{-0.1cm}
\subfigure[An iterative process for the system in Fig. 2 (a)]{
\begin{minipage}[b]{0.35\textwidth} 
\includegraphics[width=\textwidth]{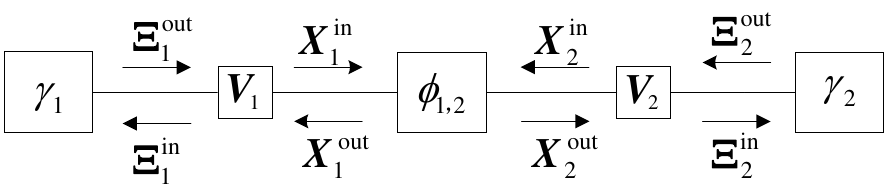} 
\end{minipage} } 
\subfigure[Local GIP at Port $1$ for the system in Fig. 2 (a)]{
\begin{minipage}[b]{0.35\textwidth} 
  \includegraphics[width=\textwidth]{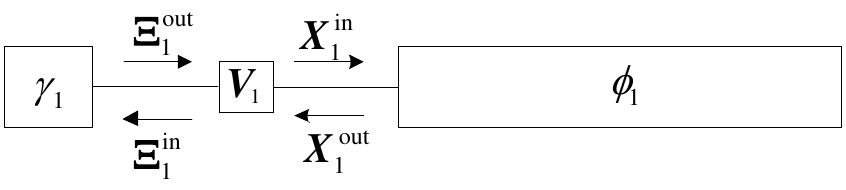}
   \end{minipage}
}
\caption{An example of an MT system.}\label{Fig:MC_network}
\end{figure}

\begin{figure}
  \centering
  \includegraphics[width=0.35\textwidth]{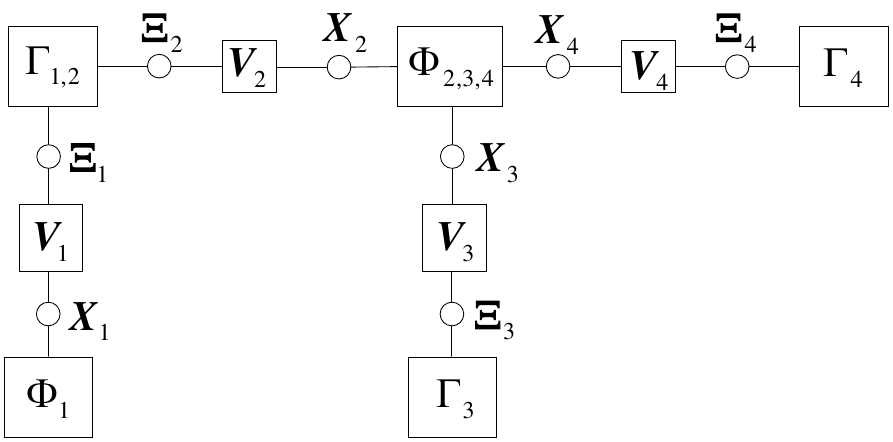}\\
   \caption{Another example of an MT system. }\label{Fig:example1}
\end{figure}

As an example, Fig. \ref{Fig:MC_network}(a) illustrates the following system: 
\BS\begin{align}  \label{Eqn:MC_example1}
 & \bf{\Xi}_1=\bf{V}_1\bf{X}_1, \;\bf{\Xi}_2=\bf{V}_2\bf{X}_2,\\
 & \bf{\Xi}_1\!\sim\! \Gamma_1, \;\bf{\Xi}_{2}\!\sim\! \Gamma_{2},\\ 
&{{\bm X}_{{{\mathcal I}_1}}} \!\!=\!\! {{\bm X}_{{{\mathcal I}_2}}}\!\!=\!\!\{\bf{X}_1,\! \bf{X}_{2}\} \!\sim\! \Phi_{1,2}. 
\end{align}\ES 
Here $\Gamma_1$ and $\Gamma_2$ are single-port constraints while $\Phi_{1,2}$ a multi-port one. The latter involves more than one input. As another example, Fig. \ref{Fig:example1} illustrates the following system:
\BS\begin{align} \label{Eqn: MC_example2}
 & \bf{\Xi}_1\!=\!\bf{V}_1\bf{X}_1, \;\bf{\Xi}_2\!=\!\bf{V}_2\bf{X}_2, \;\bf{\Xi}_3\!=\!\bf{V}_3\bf{X}_3,\;\bf{\Xi}_4\!=\!\bf{V}_4\bf{X}_4,\\
  &  
  \{\bf{\Xi}_{1}, \bf{\Xi}_{2}\} \!\sim\! \Gamma_{1,2},\;\bf{\Xi}_{3}\!\sim\! \Gamma_{3},\; \bf{\Xi}_{4}\!\sim\! \Gamma_{4},  \\
  & \bf{X}_1\!\sim\! \Phi_{1},\;  
  \{\bf{X}_2,\bf{X}_{3},\bf{X}_4\} \!\sim\! \Phi_{2,3,4}.
\end{align} \ES 
\subsection{Generic Iterative Process with MT}
Using the above notations, we can express a general MT system as follows:
\BE \label{Eqn:MC_constraints}
\!\!\!{{\bm \Xi} _k} = {{\bm V}_k}{{\bm X}_k},{{\bm \Xi} _{{{\mathcal J}_k}}} \sim {\Gamma _{{{\mathcal J}_k}}},{{\bm X}_{{{\mathcal I}_k}}} \sim {\Phi _{{{\mathcal I}_k}}},\; k=1,2,\!\cdots\!,K.
\EE
From \eqref{Eqn:Ik_example1}-\eqref{Eqn:Ik_example3} that ${\mathcal I}_k={\mathcal I}_{k'}$, $\forall k'\in {\mathcal I}_k$, and ${\mathcal J}_k={\mathcal J}_{k'}$, $\forall k'\in {\mathcal J}_k$. Hence, \eqref{Eqn:MC_constraints} may contain repeated (and so redundant) constraints. Clearly, \eqref{Eqn:MC_constraints} is a generalization of \eqref{Eqn:X_Vx_MMV}.

Based on \eqref{Eqn:MC_constraints}, we define a general iterative process as follows, which is a generalization of \eqref{Eqn:Ite_Proc_para}.

\noindent\textbf{Generic Iterative Process for an MT Problem:} Starting with $t=1$ and $ {\bm {\Xi}}_{{\mathcal J }_k}^{{\rm{in}},0} = {\bm X}_{{\mathcal I }_k}^{{\rm{in}},0} = {\bm 0}^{N\times M}$ for $k=1,2,\cdots,K$: 
\begin{subequations} \label{Eqn:Ite_GIPMC}
\begin{align}
\!\!\!{\bm {\Xi}}_{{{\mathcal J}_k }}^{{\rm{out}},t} \!\!&=\! {\gamma _{{\mathcal J }_k}^t}( { {\bm {\Xi}}_{{\mathcal J }_k}^{{\rm{in}},t-1}} ), \!&&{\bm X}_{{\mathcal I}_k}^{{\rm{out}},t} \!\!=\! {\phi _{{\mathcal I }_k}^t}( {{\bm X}_{{\mathcal I }_k}^{{\rm{in}},t-1}} ), \label{Eqn:Ite_GIPMC1}\\
   \!\!\! {\bm {\Xi}}_{k}^{{\rm{in}},t} &= {{\bm V}_k} {\bm X}_{k}^{{\rm{out}},t},\!&&{\bm {X}}_{k}^{{\rm{in}},t} = {\bm V}_k^{\rm{T}} {\bm {\Xi}}_{k}^{{\rm{out}},t}. \label{Eqn:Ite_GIPMC2}
\end{align}
\end{subequations}
In \eqref{Eqn:Ite_GIPMC}, $\phi_{{\mathcal I}_k}^t(\cdot)$ is a local processor based on $\Phi_{{\mathcal I}_k}$. Its output $ {\bm X}_{{{\mathcal I}_k}}^{{\rm{out}},t}$ is hopefully a refined estimate of ${\bm X}_{{\mathcal I}_k}$, relative to its input $ {\bm X}_{{{\mathcal I}_k}}^{{\rm{in}},t-1}$. Similar to \eqref{Eqn:Ik_example2}, we can decompose the input and output of a multi-port function as follows
\BS \label{Eqn:decompose}
\begin{align} 
{\bm X}_{{{\mathcal I}_k}}^{{\rm{in}},t-1} &= [ {{\bm X}_{{k_1}}^{{\rm{in}},t-1},{\bm X}_{{k_2}}^{{\rm{in}},t-1}, \cdots ,{\bm X}_{{k_{|{{\mathcal I}_k}|}}}^{{\rm{in}},t-1}} ],\\
{\bm X}_{{{\mathcal I}_k}}^{{\rm{out}},t} &= [ {{\bm X}_{{k_1}}^{{\rm{out}},t},{\bm X}_{{k_2}}^{{\rm{out}},t}, \cdots ,{\bm X}_{{k_{|{{\mathcal I}_k}|}}}^{{\rm{out}},t}} ],
\end{align}
\ES
where $k_1,k_2,\cdots,k_{|{{\mathcal I}_k}|}$ are the elements of ${\mathcal I}_k$, and ${{\bm X}_{{k_l}}^{{\rm{in}},t-1}}$ and ${{\bm X}_{{k_l}}^{{\rm{out}},t}}$ are those at a particular port $k_l$ of ${\mathcal I}_k$.

Similarly, we have $\gamma_{{\mathcal J}_k}^t(\cdot)$. Its input ${\bm \Xi}_{{{\mathcal J}_k}}^{{\rm{in}},t-1}$ and output ${\bm \Xi}_{{{\mathcal J}_k}}^{{\rm{out}},t}$ can be decomposed on a similar way as that in \eqref{Eqn:decompose}.

A graphic example of GIP with MT is shown in Fig. \ref{Fig:MC_network}(b) for the system in Fig. \ref{Fig:MC_network}(a). This is a straightforward extension of Fig. \ref{Fig:MMV_net}(b). Notice that Fig. \ref{Fig:MC_network}(b) involves two orthogonal transforms ${\bm V}_1$ and ${\bm V}_2$, and a multi-port local estimator $\phi_{1,2}(\cdot)$. 
As mentioned below \eqref{Eqn:MC_constraints}, there may be repeated constraints in \eqref{Eqn:MC_constraints}. This means that there can be repeated operations in \eqref{Eqn:Ite_GIPMC}. Nevertheless, \eqref{Eqn:Ite_GIPMC} is very concise, which make it easier for our discussions below. In practice, we can avoid repeated operations based on the graphic representation of \eqref{Eqn:MC_constraints}. 

\subsection{Augmented OAMP} \label{Sec:AOAMP}
We denote the GS models of the input and output in \eqref{Eqn:Ite_GIPMC} as follows. 
\begin{subequations} \label{Eqn:Ite_GIPMCEqu_GS}
\begin{align}
\!\!\!\!\! {\bm {\Xi}}_k^{{\rm{out}},t} &\!=\! {\bm {\Xi}}_k{\bm \Theta }_{{\Xi}_k}^{{\rm{out}},t}\! + \!{\bm Z }_{{ \Xi}_k}^{{\rm{out}},t}, {\bm X}_k^{{\rm{out}},t}\! =\! {\bm X}_k{\bm \Theta} _{{ X}_k}^{{\rm{out}},t} \!+\! {\bm Z }_{{ X}_k}^{{\rm{out}},t},\\
\!\!\!\!\! {\bm {\Xi}}_k^{{\rm{in}},t} &= {\bm {\Xi}}_k{\bm \Theta }_{{\Xi}_k}^{{\rm{in}},t} + {\bm Z }_{{ \Xi}_k}^{{\rm{in}},t},\; {\bm X}_k^{{\rm{in}},t} = {\bm X}_k{\bm \Theta} _{{ X}_k}^{{\rm{in}},t} + {\bm Z }_{{ X}_k}^{{\rm{in}},t}.
\end{align}
\end{subequations}
where ${\bm Z}_{{\Xi _k}}^{{\rm{out}},t}$, ${\bm Z}_{{\Xi _k}}^{{\rm{in}},t}$, ${\bm Z}_{{X _k}}^{{\rm{out}},t}$ and ${\bm Z}_{{X _k}}^{{\rm{in}},t}$ are GS errors at port $k$ in iteration $t$, with the auto-covariance matrices denoted as ${\bm \Sigma }_{{ \Xi}_k}^{{\rm{out}},t}$,${\bm \Sigma }_{{ \Xi}_k}^{{\rm{in}},t}$,${\bm \Sigma }_{{ X}_k}^{{\rm{out}},t}$ and ${\bm \Sigma }_{{ X}_k}^{{\rm{in}},t}$ respectively. 

\begin{definition}[A-OAMP]
Augmented OAMP (A-OAMP) is a special case of GIP in \eqref{Eqn:Ite_GIPMC} when the following port-wise LLN-orthogonality holds for $k=1,2,\cdots,K$, $N \to \infty $, $t \ge 1$ and $0 \le t' < t$: \begin{subequations}\label{Eqn:Orthogonality_MC}
 \begin{align}
\!\!\!\!\!\!\!{\tfrac{1}{N}( {\bm Z}_{\Xi_k} ^{{\rm{in}},t'} )^{\rm{T}}}{\bm Z}_{\Xi_k} ^{{\rm{out}},t} &\stackrel{\text{LLN}}{\longrightarrow} {\bm 0}^{M \times M},{\tfrac{1}{N}( {\bm Z}_{X_k} ^{{\rm{in}},t'} )^{\rm{T}}}{\bm Z}_{X_k} ^{{\rm{out}},t} \stackrel{\text{LLN}}{\longrightarrow} {\bm 0}^{M\times M},\\
\!\!\!\!\!\!\!\!\!\!\!{\tfrac{1}{N}{\bm {\Xi}_k^{\rm{T}}}}{\bm Z}_{\Xi_k} ^{{\rm{out}},t} &\!\stackrel{\text{LLN}}{\longrightarrow}\! {\bm 0}^{M \times M},{\tfrac{1}{N}{\bm X}_k^{\rm{T}}}{\bm Z}_{X_k} ^{{\rm{out}},t} \!\stackrel{\text{LLN}}{\longrightarrow}\! {\bm 0}^{M \times M}.
\end{align} 
 \end{subequations}
\end{definition}
In \eqref{Eqn:Orthogonality_MC}, the input and output errors at each port are required to be orthogonal to each other. There is no orthogonality requirement across different ports. It is seen that \eqref{Eqn:Orthogonality} is a special case of \eqref{Eqn:Orthogonality_MC} when $K=1$. 

If all ${\bm X}_k$ in \eqref{Eqn:Orthogonality_MC} are of single column, OAMP-MT is a proper name for the related algorithm. However, if some of ${\bm X}_k$ in \eqref{Eqn:Orthogonality_MC} contain multiple columns, OAMP-MMV developed in Sec. III are required in the realization of \eqref{Eqn:Orthogonality_MC}. In this case, OAMP-MMV-MT can be a more appropriate name, but this reads somewhat cumbersome. Hence, the name A-OAMP is used instead.

\subsection{Equivalent ST System and Its Error Behaviour} \label{Sec:Mesh}
We now extend Theorem~\ref{Theo:IIDG_MMV} from ST to MT. 

Recall the discussions related to \eqref{Eqn:block-wise}. For $\gamma_{{\mathcal J}_k}^t(\cdot)$ and $\phi_{{\mathcal I}_k}^t(\cdot)$, we define their port-wise function at port $k$:
\begin{equation} \label{Eqn:Port_phi}
{\bm \Xi} _k^{{\rm{out}},t} = \gamma _k^t( {{\bm \Xi} _k^{{\rm{in}},t - 1}} ),\;\;\; {\bm X} _k^{{\rm{out}},t} = \phi_k^t ( {{\bm X} _k^{{\rm{in}},t - 1}} ).
\end{equation}
Keep in mind that the inputs into $\gamma_k^t(\cdot)$ and $\phi_k^t(\cdot)$ (i.e., the inputs from other ports $k' \ne k$) that do not appear in \eqref{Eqn:Port_phi} are all treated as parameters. Using \eqref{Eqn:Ite_GIPMC} and \eqref{Eqn:Port_phi}, we can write a local GIP at port $k$ as follows.

\noindent\textbf{Local GIP at Port $k$:} Starting with $t=1$, and $ {\bm {\Xi}}_{k}^{{\rm{in}},0} =  {\bm X}_{k}^{{\rm{in}},0}= {\bm 0}^{N \times M}$: 
\begin{subequations} \label{Eqn:Ite_GIPMC_Equiv}
\begin{align}
     {\bm {\Xi}}_{k}^{{\rm{out}},t} &= {\gamma _{k}^t}( { {\bm {\Xi}}_{k}^{{\rm{in}},t-1}} ),&& {\bm X}_{k}^{{\rm{out}},t} = {\phi _{k}^t}( { {\bm X}_{k}^{{\rm{in}},t-1}} ),\\
 {\bm {\Xi}}_{k}^{{\rm{in}},t} &= {{\bm V}_k}{\bm X}_{k}^{{\rm{out}},t},&& {\bm X}_{k}^{{\rm{in}},t} ={\bm V}_k^{\rm{T}} {\bm {\Xi}}_{k}^{{\rm{out}},t}.
\end{align}
\end{subequations}

For example, Fig. \ref{Fig:MC_network}(c) illustrates the local GIP related to ${\bm V}_1$ in Fig. \ref{Fig:MC_network}(b), where $\phi_1(\cdot)$ is the port-wise function of $\phi_{1,2}(\cdot)$ at port 1. In iteration $t$, the output of $\phi_1^t(\cdot)$ is determined by its input ${\bm X}_1^{{\rm{in}},t-1}$ together with its implicit parameters $\{{\bm X}_2^{{\rm{in}},t-1}, {\bm X}_2^{{\rm{out}},t}\}$, ${\bm V}_2$ and $\{{\bm \Xi}_2^{{\rm{in}},t-1}, {\bm \Xi}_2^{{\rm{out}},t}\}$.  This can be seen by comparing Fig. \ref{Fig:MC_network}(b) and Fig. \ref{Fig:MC_network}(c). This example indicates the following.

\begin{remark}
In general, $\gamma_k(\cdot)$ and $\phi_k(\cdot)$ can be functions of ${\bm V}_k', k'=1,2,\cdots,K, k'\ne k$.
\end{remark}

Except for the internal structures of $\gamma_k(\cdot)$ and $\phi_k(\cdot)$, the resemblance between \eqref{Eqn:Ite_GIPMC_Equiv} and \eqref{Eqn:Ite_Proc_para} is apparent. Due to this, Theorem~\ref{Theo:IIDG_MMV} can be applied to \eqref{Eqn:Ite_GIPMC_Equiv} provided that Assumption~\ref{Asu:MMV} holds for \eqref{Eqn:Ite_GIPMC_Equiv}. To this end, we introduce the assumptions below for the GIP in \eqref{Eqn:Ite_GIPMC}.

\begin{assumption} \label{Ass:MC}
(i) At the beginning of each experiment, every $\bf{V}_{k}$ is independently sampled using ${{\bm V}_k} \sim {\bm{\mathcal{H}}^{N \times N}}$, so $\{ {{{\bm V}_1},{{\bm V}_2}, \cdots ,{{\bm V}_K}} \}$ are mutually independent Haar distributed matrices. (ii) During each experiment, every ${\bm V}_k$ remains unchanged for all iterations. (iii) Both $\{\phi_{{\mathcal I}_k}^t(\cdot)\}$ and $\{\gamma_{{\mathcal J}_k}^t(\cdot)\}$ are PIID-separable and independent of ${\bm V}_k$, $k=1,2,\cdots,K$.
\end{assumption}

\underline{\textbf{Notes:}} Clearly, items (i) and (ii) in Assumption~\ref{Ass:MC} ensure items (i) and (ii) in Assumption~\ref{Asu:MMV}. In the proof of Theorem~\ref{Theo:IIDG_MC} below, the key is to prove item (iii) in Assumption~\ref{Asu:MMV} holds under Assumption~\ref{Ass:MC}.

 \begin{theorem} \label{Theo:IIDG_MC}
Under Assumption \ref{Ass:MC}, let A-OAMP be initialized with ${{\bm {\Xi}}}_{k}^{{\rm{in}},0} = {\bm { X}}_{k}^{{\rm{in}},0} = {\bm 0}^{N \times M}$, $k=1,2,\cdots,K$. Then the following claims hold for any finite $t$ for $N \to \infty$: 
 \begin{itemize}
  \item[(i)] for any $k$, ${\bm Z}_{\Xi_k} ^{{\rm{in}},\{ t-1 \}}$ and ${\bm Z}_{X_k} ^{{\rm{in}},\{ t-1 \}}$ converge to CPIIDG-RJG in distribution;
    \item[(ii)] for any $k$, the angles of ${\bm Z}_{\Xi_k} ^{{\rm{in}},\{ t-1 \}}$ and ${\bm Z}_{X_k} ^{{\rm{in}},\{ t-1 \}}$ are asymptotically entry-wise independent of ${\bm z}_k$ in the Markov chain below:
\vspace{-0.1cm} \BE 
  \!\!\!\!\!  {\bm z}_k \!\to\! [{\bm Z}_{\Xi_k}^{{\rm{out}},t}\!,\!{\bm Z}_{X_k}^{{\rm{out}},t}\!,\!{{\bm A}_k^t,\!{\bm B}_k^t}] \!\to\! {\bm V}_k \!\sim\! {\bm {\mathcal H}}(\!{ {{\bm A}_k^t \!=\! {\bm V}_k{\bm B}_k^t} }\!) \nonumber. \vspace{-0.1cm}
   \EE
\end{itemize}  
\end{theorem} 
\begin{IEEEproof}
We prove by induction on $t$. At $t=0$, similar to Theorem~\ref{Theo:IIDG_MMV}, zero error matrices can be treated as special cases of CPIIDG-RJG and asymptotically independent of ${\bm z}_k$. Hence Theorem~\ref{Theo:IIDG_MC} holds.

Now assume that Theorem~\ref{Theo:IIDG_MC} holds for any $t > 0$ and consider $t+1$. As mentioned above, (i) and (ii) in Assumption~\ref{Ass:MC} ensure (i) and (ii) in Assumption~\ref{Asu:MMV}. Note that (a) each ${\bm Z}_{{X_{{{\mathcal I}_k}}}}^{{\rm{in}},t-1}$ can be decomposed as $\{ {{\bm Z}_{{X_{{k_1}}}}^{{\rm{in}},t - 1},{\bm Z}_{{X_{{k_2}}}}^{{\rm{in}},t - 1}, \cdots ,{\bm Z}_{{X_{{k_{|{{\mathcal I}_k}|}}}}}^{{\rm{in}},t - 1}} \}$ (see \eqref{Eqn:decompose}), and (b) from induction, ${{\bm Z}_{{X_{{k}}}}^{{\rm{in}},t - 1}}, k=1,2,\cdots,K$ are all CPIIDG-RJG. Combining (a) and (b) and Lemma~\ref{Lem:CPIID_2}, we can see that ${\bm Z}_{{X_{{{\mathcal I}_k}}}}^{{\rm{in}},t-1}$, $k=1,2,\cdots,K$ are all CPIIDG-RJG. Then from Assumption~\ref{Ass:MC}(iii) and Lemma~\ref{Lem:multi-port-PIID}, all $\{\phi_k^t(\cdot)\}$ are PIID-separable. Similarly, we can show that all $\{\gamma_k^t(\cdot)\}$ are also PIID-separable. Assumption~\ref{Ass:MC} ensures that $\phi_k^t(\cdot)$ and $\gamma_k^t(\cdot)$ are independent of ${\bm V}_k$, even though $\phi_k^t(\cdot)$ and $\gamma_k^t(\cdot)$ can be functions of ${\bm V}_k', k'\ne k$. Hence Assumption~\ref{Asu:MMV}(iii) also holds for \eqref{Eqn:Ite_GIPMC_Equiv}. Then following Theorem~\ref{Theo:IIDG_MMV}, Theorem~\ref{Theo:IIDG_MC} holds for $t+1$. This completes the induction.  
\end{IEEEproof}
\subsection{State Evolution of A-OAMP} \label{Sec:MC_SE}
Denote by ${\rm{diag}}\{{\mathbb S }\}$ a diagonal matrix whose entries are the elements in a set of ${\mathbb S}$. The block-diagonal entries in the matrices below are the port-wise GS parameters given in \eqref{Eqn:Ite_GIPMCEqu_GS}.
\begin{subequations} \label{Eqn:GIPMC_GSpara}
\begin{align}
 {\bm \Theta} _{{\Xi _{{\mathcal J}_k}}}^{{\rm{in}},t} &= {\rm{diag}}\{ {{\bm \Theta} _{{\Xi _{k'}}}^{{\rm{in}},t},k' \in {{\mathcal J}_k }} \},\\
{\bm \Theta} _{{\Xi _{{\mathcal J}_k}}}^{{\rm{out}},t} &= {\rm{diag}}\{ {{\bm \Theta} _{{\Xi _{k'}}}^{{\rm{out}},t},k' \in {{\mathcal J}_k}} \},\\
{\bm \Theta} _{{X _{{\mathcal I}_k}}}^{{\rm{in}},t} &= {\rm{diag}}\{ {{\bm \Theta} _{{X _{k'}}}^{{\rm{in}},t},k' \in {{\mathcal I}_k }} \},\\
{\bm \Theta} _{{X _{{\mathcal I}_k}}}^{{\rm{out}},t} &= {\rm{diag}}\{ {{\bm \Theta} _{{X _{k'}}}^{{\rm{out}},t},k' \in {{\mathcal I}_k }} \}, 
\end{align}
\end{subequations}
and ${\bm \Sigma}_{{\Xi _{{\mathcal J}_k}}}^{{\rm{in}},t}, {\bm \Sigma} _{{\Xi _{{\mathcal J}_k}}}^{{\rm{out}},t}, {\bm \Sigma} _{{X _{{\mathcal I}_k}}}^{{\rm{in}},t}$, ${\bm \Sigma} _{{X _{{\mathcal I}_k}}}^{{\rm{out}},t}$ can be defined similarly. We can rewrite \eqref{Eqn:Ite_GIPMCEqu_GS} in a matrix form as follows:
\begin{subequations} \label{Eqn:Ite_GIPMC_GS}
\begin{align}
 {\bm {\Xi}}_{{\mathcal J}_k}^{{\rm{in}},t} &= {{\bm {\Xi}}_{{\mathcal J}_k}}{\bm \Theta} _{{\Xi _{{\mathcal J}_k}}}^{{\rm{in}},t} + {\bm Z}_{{\Xi _{{{\mathcal J}_k}}}}^{{\rm{in}},t},\\
 {\bm {\Xi}}_{{\mathcal J}_k}^{{\rm{out}},t} &= {{\bm {\Xi}}_{{\mathcal J}_k}}{\bm \Theta} _{{\Xi _{{\mathcal J}_k}}}^{{\rm{out}},t} + {\bm Z}_{{\Xi _{{{\mathcal J}_k}}}}^{{\rm{out}},t},\\
      {\bm X}_{{\mathcal I}_k}^{{\rm{in}},t} &= {\bm X}_{{\mathcal I}_k}{\bm \Theta }_{{X _{{\mathcal I}_k}}}^{{\rm{in}},t} + {\bm Z}_{{X _{{{\mathcal I}_k}}}}^{{\rm{in}},t},\\
    {\bm X}_{{\mathcal I}_k}^{{\rm{out}},t} &= {\bm X}_{{\mathcal I}_k}{\bm \Theta }_{{X _{{\mathcal I}_k}}}^{{\rm{out}},t} + {\bm Z}_{{X _{{{\mathcal I}_k}}}}^{{\rm{out}},t}.
    \end{align}
\end{subequations}
The difference between \eqref{Eqn:Ite_GIPMCEqu_GS} and \eqref{Eqn:Ite_GIPMC_GS} is that the former is for a particular port $k$ while the latter for all the ports of each multi-port estimator.   

The following SE recursions characterize the GS parameters in \eqref{Eqn:GIPMC_GSpara} for A-OAMP. 

\underline{State evolution for A-OAMP:} Starting with $t=1$, ${\bm \Theta} _{{\Xi _{{\mathcal J}_k}}}^{{\rm{in}},0}\!\! = \!\!{\bm \Theta} _{{X _{{\mathcal I}_k}}}^{{\rm{in}},0}\!\!=\!\!{\bm 0}^{M\times M}$ and ${\bm \Sigma} _{{\Xi _{{\mathcal J}_k}}}^{{\rm{in}},0}\!\! = \!\!{\bm \Sigma} _{{X _{{\mathcal I}_k}}}^{{\rm{in}},0}\!\!=\!\!{\bm 0}^{M\times M}$ for $\forall  k  $:
\begin{subequations} \label{Eqn:GIPMC_SE}
\begin{align}
  \!\!  ( {\bm \Theta} _{{\Xi _{{\mathcal J}_k}}}^{{\rm{out}},t}, {\bm \Sigma} _{{\Xi _{{\mathcal J}_k}}}^{{\rm{out}},t} ) &\!=\! \gamma _{_{{\rm{SE}}},{\mathcal J}_k}^t( {\bm \Theta} _{{\Xi _{{\mathcal J}_k}}}^{{\rm{in}},t-1},{\bm \Sigma} _{{\Xi _{{\mathcal J}_k}}}^{{\rm{in}},t-1} ),\\
\!\!( {\bm \Theta} _{{X _{{\mathcal I}_k}}}^{{\rm{out}},t},{\bm \Sigma} _{{X _{{\mathcal I}_k}}}^{{\rm{out}},t}) &\!=\! \phi _{{{\rm{SE}}},{\mathcal I }_k}^t( {\bm \Theta} _{{X _{{\mathcal I}_k}}}^{{\rm{in}},t-1},{\bm \Sigma} _{{X_{{\mathcal I}_k}}}^{{\rm{in}},t-1}),\\
\!\!\{   {\bm \Theta} _{{\Xi _{{\mathcal J}_k}}}^{{\rm{out}},t}, {\bm \Sigma} _{{\Xi _{{\mathcal J}_k}}}^{{\rm{out}},t} \} &\stackrel{\text{re-group}}{\longrightarrow} \{ {\bm \Theta} _{{X _{{\mathcal I}_k}}}^{{\rm{in}},t}, {\bm \Sigma} _{{X _{{\mathcal I}_k}}}^{{\rm{in}},t}\}, \label{Eqn:GIPMC_SEc}\\
\!\!\{ {\bm \Theta} _{{X _{{\mathcal I}_k}}}^{{\rm{out}},t}, {\bm \Sigma} _{{X _{{\mathcal I}_k}}}^{{\rm{out}},t} \} &\stackrel{\text{re-group}}{\longrightarrow} \{ {\bm \Theta} _{{\Xi _{{\mathcal J}_k}}}^{{\rm{in}},t}, {\bm \Sigma} _{{\Xi _{{\mathcal J}_k}}}^{{\rm{in}},t}\}. \label{Eqn:GIPMC_SEd}
\end{align}
\end{subequations}

{\underline{{\textbf{Notes}}:} Under Theorem \ref{Theo:IIDG_MC}, we can model the input errors of ${\gamma _{{\mathcal J }_k}^t}(\cdot)$ and ${\phi _{{\mathcal I }_k}^t}(\cdot)$ with PIIDG variables with the same mean and variances. We can then generate $\gamma _{{{\rm{SE}}},{\mathcal J }_k}^t(\cdot)$ and $\phi _{{{\rm{SE}}},{\mathcal I }_k}^t(\cdot)$ in \eqref{Eqn:GIPMC_SE} either by analysis or simulations.

\subsection{GSO for a Multi-Port Estimator} \label{Sec:MC_GSO}
For a multi-port estimator ${\phi _{{\mathcal I }_k}^t}(  \cdot  )$, we can apply GSO in \eqref{Eqn:GGSO_B} to each port as follows:
\begin{equation} \label{Eqn:GSOsingleport}
    \!\!\!\! {\bm X}_{k}^{{\rm{out}},t} = {\phi _{k}^t}( { {\bm X}_{k}^{{\rm{in}},t-1}} )\! =\! {{\hat \phi }_{k}^t}( {{\bm X}_{k}^{{\rm{in}},{t-1}}} ) \!-\!  {\bm X}_{k}^{{\rm{in}},t-1}{{\bm \Delta}_{k}^t}, \forall k 
\end{equation}
with 
\begin{equation} \label{Eqn:GSO_B}
\bf{\Delta}_{k}^t \!=\! [\mr{E}\{({\bm Z}_{{X_k}}^{{\rm{in}},t-1})^{\mr{T}}{\bm Z}_{{X_k}}^{{\rm{in}},t-1}\}]^{-1}\mr{E}\{ ({\bm Z}_{{X_k}}^{{\rm{in}},t-1})^{\mr{T}}\hat{\phi}_{k}^t( {\bm X}_k^{{\rm{in}},t - 1})\}. 
\end{equation}

In practice, we can orthogonalize a multi-port function ${\phi _{{\mathcal I }_k}^t}(  \cdot  )$ as follows. First, we can replace the GS errors in the GS models of the inputs to ${{\hat \phi} _{{\mathcal I }_k}^t}(  \cdot  )$ using independent additive Gaussian noise samples. Second, we carry out Monte Carlo simulations to obtain the port-wise correlation $\mr{E}\{ ({\bm Z}_{{X_k}}^{{\rm{in}},t-1})^{\mr{T}}\hat{\phi}_{k}^t( {\bm X}_k^{{\rm{in}},t - 1})\}$. Third, we orthognalize each port $k$ using GSO in \eqref{Eqn:GSOsingleport}. 

Similarly, we can orthogonalize a prototype ${\gamma _{{\mathcal J}_k}^t}(\cdot)$ in a different domain via GSO. 

The proof of the following theorem is omitted since it is similar to the proof of Theorem \ref{Theo:GSO_OAMP_MMV}.

\begin{theorem}\label{Theo:GSO_OAMP}
Assume that $\{\gamma_{{\mathcal I}_k}^t(\cdot)\}$ and $\{\phi_{{\mathcal J}_k}^t(\cdot)\}$ constructed using GSO in Sec.~\ref{Sec:MC_GSO} at iterations from $0$ to $t$ for A-OAMP are orthogonal. Then the port-wise LLN-orthogonality in \eqref{Eqn:Orthogonality_MC} holds for iterations from $0$ to $t$.  
\end{theorem}

The above provides a realization technique for A-OAMP in MMV and MT problems.

\section{Application: MIMO-relay System with Multiple Source Data Streams}\label{Sec:sectioned_CQ}
In this section, we consider an MIMO relay system with multiple correlated source data \cite{relay2008,relay2011}. In such an MIMO system, the received signal at the relay node can have a high peak to average power ratio (PAPR). Clipping is used at the relay node to reduce the PAPR and improve amplifier power efficiency \cite{Han2005}. The overall system can be modelled as a joint MMV-MT problem. A-OAMP will be used to treat the non-linear effect of clipping. 

Similar techniques may also be used to treat other systems, such as orthogonal frequency division multiplexing (OFDM), with similar high PAPR problem.

\subsection{System Overview}
Fig. \ref{Fig:mesh_CQ}(a) illustrates an MIMO-relay system consisting of three nodes \cite{relay2008}, where the source, relay and
destination nodes are equipped with $N_s$,$N_r$ and $N_d$ antennas respectively. The system is modelled as follows.
\begin{itemize}
\item In the first phase, the source node transmits modulated signal ${\bm X}_s$ to the relay node over an ${N_r \!\times\! N_s}$ MIMO channel ${\bm H}_{sr}$. In the second phase, the relay node performs clipping on the received signal and then transmits it to the destination over an $N_d \!\times\! N_r$ MIMO channel ${\bm H}_{rd}$. 
\item There is no direct path between the source node and the destination.
\item Memoryless and Quasi-static frequency-flat fading channel is assumed.
\end{itemize}
\begin{figure*}[b!]
  \centering
  \includegraphics[width=0.75\textwidth]{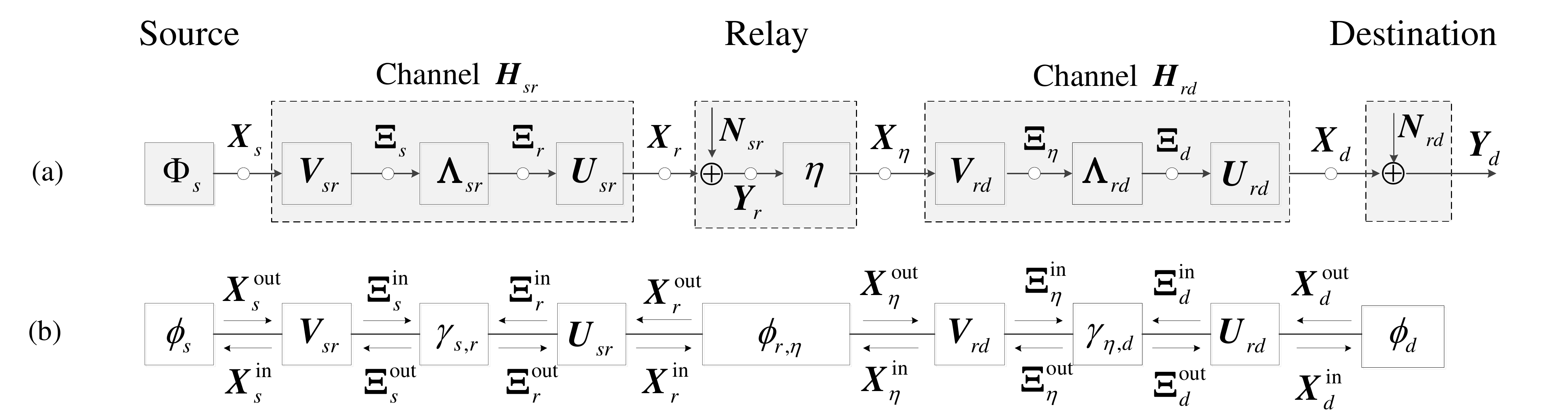}\\ \vspace{-0.15cm}
  \caption{Graphic illustrations for (a) system model of an MIMO-relay network with clipping, as expressed in \eqref{Eqn:mesh_CQ} and \eqref{Eqn:mesh_CQ_graphic}, and (b) the iterative process for the system illustrated in Fig. \ref{Fig:mesh_CQ}(a), where $\phi_s$, $\gamma_{s,r}$, $\phi_{r,\eta}$, $\gamma_{\eta,d}$ and $\phi_d$ are corresponding estimators for the constraints in Fig. \ref{Fig:mesh_CQ}(a).}\label{Fig:mesh_CQ}
\end{figure*} 

Assume that the transmitted signal ${\bm X}_s\in\mathbb{R}^{N_s\times M}$ is binary phase shift keying (BPSK) modulated. The received signal at the relay and destination nodes are respectively given by
\BS \label{Eqn:received_relay}
\begin{align}
{\bm Y}_r&={\bm H}_{sr}{\bm X}_s+{\bm N}_{sr},\\
{{\bm Y}_d} &= {{\bm H}_{rd}} \cdot \eta ( {{{\bm Y}_r}} ) + {{\bm N}_{rd}},\label{Eqn:received_relayb}
\end{align}
\ES
where ${\bm H}_{sr} \!\!\in\!\! \mathbb{C}^{N_r \times N_s}$ and
${\bm H}_{rd} \!\!\in\!\! \mathbb{C}^{N_d \times N_r}$ are respectively the channel matrices for the source-relay and relay-destination links, and ${\bm N}_{sr} $ and ${\bm N}_{rd}$ are two matrices of IID additive noise samples following Gaussian distribution 
$\mathcal{CN}(0,v_{sr})$ and $\mathcal{CN}(0,v_{rd})$ respectively. The details on the clipping function $\eta(\cdot)$ are given below. 

By the central limit theorem, the entries of ${\bm Y}_{r}$ are approximately Gaussian distributed and thus have a high PAPR. Directly transmitting ${\bm Y}_r$ from the delay would require stringent RF amplifier linearity, which is difficult in practice. To reduce PAPR \cite{Han2005}, we adopt a symbol-by-symbol clipping function at the real part as follows:
\BE 
\!\!\!\!\!\!{\mr{clip}}\left( {\rm Re} ( [{\bm Y}_{r}]_i) \right) \!=\!\left\{\!\! \begin{array}{l}
Z, \quad\quad\quad\quad\; \;\;{\rm Re} ( [{\bm Y}_{r}]_i)>Z,\\
{\rm Re} ([ {\bm Y}_{r}]_i), \quad -Z\leq {\rm Re} ( [{\bm Y}_{r}]_i) \leq Z,\\
-Z, \quad\quad\quad\quad\;{\rm Re} ([ {\bm Y}_{r}]_i)<-Z.
\end{array} \right.
\EE
where $i=1,\cdots,N_r$. Similarly we can define ${\mr{clip}}\left( {\rm Im} ( {\bm Y}_{r}) \right)$. Then 
\[{\rm{clip}}( {{{\bm Y}_{r}}} ) = {\mr{clip}}\left( {\rm Re} ( {\bm Y}_{r}) \right)+ \sqrt { - 1} \cdot {\mr{clip}}\left( {\rm Im} ( {\bm Y}_{r}) \right).\]
Given the threshold $Z$, the clipping ratio (CR) is defined as ${\rm{CR \!=\! 10lo}}{{\rm{g}}_{10}}( {{Z^2}/{\rm{E}}\{ {{{| {{Y_r}} |}^2}} \}} )$. 

The transmitted signal from the relay node to the destination is then given by a normalized clipping function defined as:
\BE
\eta ( {{{\bm Y}_{r}}} ) = {\rm{clip}}( {{{\bm Y}_{r}}} )/C,
\EE
where $C = \sqrt {{\rm E} \{|{\mr{clip}}( Y_r)|^2\}}$. 
\subsection{Details}
Let ${{\bm H}_{sr}} = {{\bm U}_{sr}}{{\bm \Lambda} _{sr}}{{\bm V}_{sr}}$ and ${{\bm H}_{rd}} = {{\bm U}_{rd}}{{\bm \Lambda} _{rd}}{{\bm V}_{rd}}$. The system in \eqref{Eqn:received_relayb} can be rewritten as
\BS\label{Eqn:mesh_CQ} \begin{align}
{{\bm Y}_d} &={{\bm H}_{rd}}\cdot\eta ( {\bm Y}_r ) + {{\bm N}_{rd}},\\
&= {{\bm H}_{rd}}\cdot\eta ( {{{\bm H}_{sr}}{\bm X}_s + {{\bm N}_{sr}}} ) + {{\bm N}_{rd}},\\
&= {{\bm U}_{rd}}{{\bm \Lambda} _{rd}}{{\bm V}_{rd}}\cdot\eta ( {{{\bm U}_{sr}}{{\bm \Lambda} _{sr}}{{\bm V}_{sr}}{\bm X}_s + {{\bm N}_{sr}}} ) + {{\bm N}_{rd}}.
\end{align}
\ES
Our task is to estimate ${\bm X}_s$ based on ${\bm Y}_d$ in \eqref{Eqn:mesh_CQ}. 

Define ${{\bm X}_r} = {{\bm H}_{sr}}{{\bm X}_s}$, ${{\bm X}_\eta } = \eta ( {{{\bm Y}_r}} )$ and ${{\bm X}_d} = {{\bm H}_{rd}}{{\bm X}_{\eta}}$. Then the above system can be expressed as below, with its illustration given in Fig. \ref{Fig:mesh_CQ}(a):
\begin{subequations} \label{Eqn:mesh_CQ_graphic}
\begin{align}
&{{\bm \Xi} _s} = {{\bm V}_{sr}}{{\bm X}_s},\;\; {{\bm \Xi} _r} = {\bm U}_{sr}^{\rm{H}}{{\bm X}_r},\;\; {{\bm \Xi} _{\eta}} = {{\bm V}_{rd}}{{\bm X}_{\eta}},\\
&{{\bm \Xi} _d} = {\bm U}_{rd}^{\rm{H}}{{\bm X}_d},\;\quad\quad\quad{\Phi _s}:{{\bm X}_s} \sim p( {\bm X}_s ),\\
&{\Gamma _{s,r}}:{{\bm \Xi} _r} = {{\bm \Lambda} _{sr}}{{\bm \Xi} _s},\quad {\Phi _{r,{\eta}}}:{{\bm X}_{\eta}} = \eta ( {{{\bm X}_r} + {{\bm N}_{sr}}} ),\\
&{\Gamma _{{\eta},d}}:{{\bm \Xi} _d} = {{\bm \Lambda} _{rd}}{{\bm \Xi} _{\eta}},\quad {\Phi _d}:{{\bm Y}_d} = {{\bm X}_d} + {{\bm N}_{rd}}.
\end{align}
\end{subequations}

We can design prototype estimators corresponding to the above constraints respectively, and then realize port-wise orthogonality using \eqref{Eqn:GSOsingleport}. A graphic example of the iterative process for the above system is illustrated in Fig. \ref{Fig:mesh_CQ}(b), where $\phi_s$, $\gamma_{s,r}$, $\phi_{r,\eta}$, $\gamma_{\eta,d}$ and $\phi_d$ are corresponding estimators for $\Phi_s$, $\Gamma_{s,r}$, $\Phi_{r,\eta}$, $\Gamma_{\eta,d}$ and $\Phi_d$ respectively. 
\begin{itemize}
\item The optimal prototype estimators are usually designed based on the MMSE principle. For more details of designing local estimators, see \cite{Yiyaonotes}.
\item To simplify derivation and reduce complexity, if ${\bm X}_s$ is a matrix (i.e., $M\!>\!1$), we adopt a sub-optimal prototype estimator ${\hat \phi}_{r,\eta}$ where ${\hat \phi}_{r,\eta}(\cdot)$ is applied to each column pair $({[ {{\bm X}_r^{{\rm{in}}}} ]_{|m}},{[ {{\bm X}_{\eta}^{{\rm{in}}}} ]_{|m}})$ separately.
\item Generally speaking, we can use Monte-Carlo simulation to calculate the GSO parameters in \eqref{Eqn:GSO_B}. A special case is for the linear estimator where the GSO parameters can be explicitly given [(35),\cite{Yiyao2020}].

\end{itemize}

The following simulation setups are assumed unless otherwise stated. To reduce the computational complexity, we generate a large random unitary matrix by randomly row-wise permuting a discrete Fourier transform (DFT) matrix \cite{Ma2019}, i.e., ${{\bm U}_{sr}} = {{\bm \Pi }_{{U_{sr}}}}{\bm F}$ and ${{\bm V}_{sr}} = {{\bm \Pi }_{{V_{sr}}}}{\bm F}$, where ${{\bm \Pi} _{{U_{sr}}}}$ and ${{\bm \Pi} _{{V_{sr}}}}$ are random permutation matrices, and ${\bm F}$ is a discrete Fourier transform
(DFT) matrix. We generate ${{\bm \Lambda} _{sr}}$ in \eqref{Eqn:mesh_CQ} as ${\lambda _i}/{\lambda _{i+1}} = {\kappa^{1/{N_r}}}$ for $i = 1, \cdots ,{N_r} - 1$ and $\sum_{i = 1}^{N_r} {{\lambda _i^2}}  = N_s$ where $\kappa$ denotes the condition number of ${\bm H_{sr}}$. ${\bm \Lambda} _{rd}$ is generated similarly.
The signal to noise ratios are defined as ${\rm{SNR}}_{sr} = 1/{v_{sr}}$ and ${\rm{SNR}}_{rd} = 1/{v_{rd}}$ respectively.
\vspace{-0.1cm}\begin{figure}[b!]
  \centering
  \includegraphics[width=0.35\textwidth]{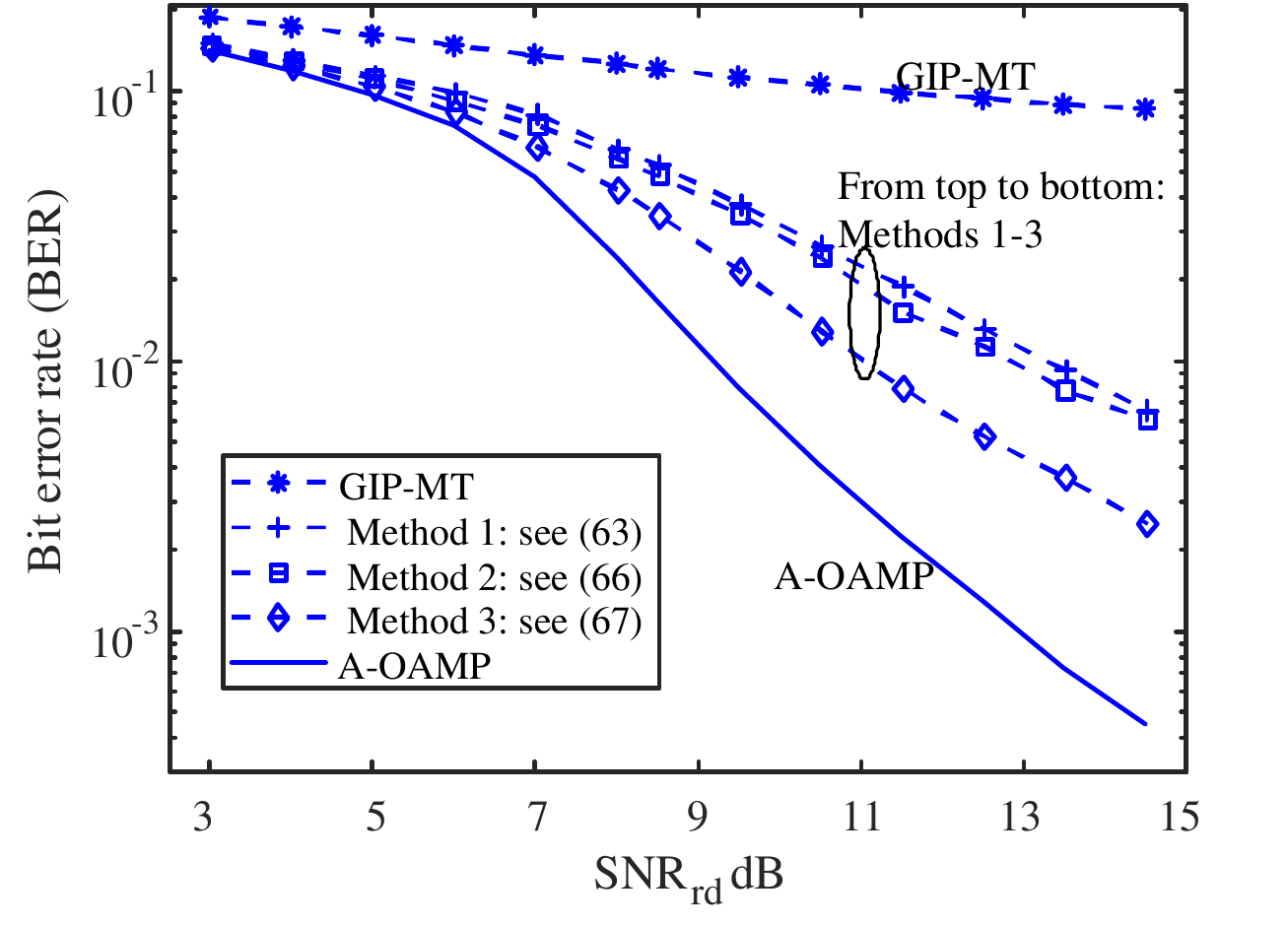}\\ \vspace{-0.15cm}
  \caption{Simulated MSEs with $SNR_{\rm{rd}}$. 
  The parameters are as follows: $N_s=8096, N_r/N_s=0.8$ and $N_d/N_r=0.8$, $SNR_{\rm{sr}}$=11dB., $\kappa_{sr}=5$, $\kappa_{rd}=5$. clipping ratios $CR =0$dB.  }\label{Fig:Qant_net}
\end{figure} 

\subsection{Existing Treatments}
To the best of our knowledge, there is no systematic discussion on the relay system in \eqref{Eqn:mesh_CQ} with clipping. Here we consider some simple treatments.

1) {\textbf{Method 1}}: We can simply ignore the clipping effect and pretend 
\BE \label{Eqn:method1}
\eta ( {{{\bm Y}_r}} ) = {{\bm Y}_r}.
\EE
In this case, \eqref{Eqn:mesh_CQ} can be simply written as
\BS \label{Eqn:mesh_CQ_equiv}
\begin{align}
{{\bm Y}_d} &= {{\bm H}_{rd}}( {{{\bm H}_{sr}}{\bm X}_s + {{\bm N}_{sr}}} ) + {{\bm N}_{rd}},\\
& = \underbrace {( {{{\bm H}_{rd}}{{\bm H}_{sr}}} )}_{{\text{Combined matrix}}}{\bm X}_s + \underbrace {( {{{\bm H}_{rd}}{{\bm N}_{sr}} + {{\bm N}_{rd}}} )}_{\text{Combined noise}}.
\end{align}
\ES
The auto-correlation of the combined noise in \eqref{Eqn:mesh_CQ_equiv} is ${{\bm \Sigma} _{{{ N}_{\rm{combined}}}}} = {v_{sr}}{{\bm H}_{rd}}{\bm H}_{rd}^{\rm{H}} + {v_{rd}}{\bm I}$. Define
\BE \label{Eqn:white_noise}
\begin{aligned}
\!\!\!\!{\tilde {\bm Y}_d}={\bm \Sigma} _{{{ N}_{{\rm{combined}}}}}^{ - 1/2}{{\bm Y}_d} &= ( {{\bm \Sigma} _{{{ N}_{{\rm{combined}}}}}^{ - 1/2}{{\bm H}_{rd}}{{\bm H}_{sr}}} ){\bm X}_s \\ &+\underbrace {{\bm \Sigma} _{{{ N}_{{\rm{combined}}}}}^{ - 1/2}( {{{\bm H}_{rd}}{{\bm N}_{sr}} + {{\bm N}_{rd}}} )}_{{\rm{white noise }}\;{N_{{\rm{white}}}}},
\end{aligned}
\EE
with each entry of ${\bm N}_{\rm{white}} \sim \mathcal{CN}(0,1)$. Then \eqref{Eqn:white_noise} can be directly solved by OAMP.

2) {\textbf{Method 2}}: An alternative way is to model  the clipped signal using an additive error noise ${\bm N}_{\eta}$ as,
\BE \label{Eqn:method2}
\eta ( {{{\bm Y}_r}} ) = {{\bm Y}_r} + {{\bm N}_\eta }.
\EE
We solve the system in \eqref{Eqn:method2} by assuming that ${\bm N}_{\eta}$ is IIDG, which is not true in practice and so results in performance loss.  

3) {\textbf{Method 3}}: We can also model the clipping effect by a GS model, i.e., 
\BE \label{Eqn:method3}
\eta ( {{{\bm Y}_r}} )=  \bf{Y}_{r}{\bm \Theta}_{\eta} + {\bm F}_{\eta},
\EE
where ${\bm \Theta}_{\eta} $ is a matrix with size $M\times M$ and ${\bm F}_{\eta}$ is the distortion uncorrelated with ${\bf{Y}}_r$. Again we assume that ${\bm F}_{\eta}$ is IIDG, which is also not true in practice. Method 3 outperforms method 2 slightly due to the orthogonality between ${\bm Y}_r$ and ${\bm F}_{\eta}$ in \eqref{Eqn:method3}.
 \begin{figure}[t] 
  \centering
  \includegraphics[width=0.35\textwidth]{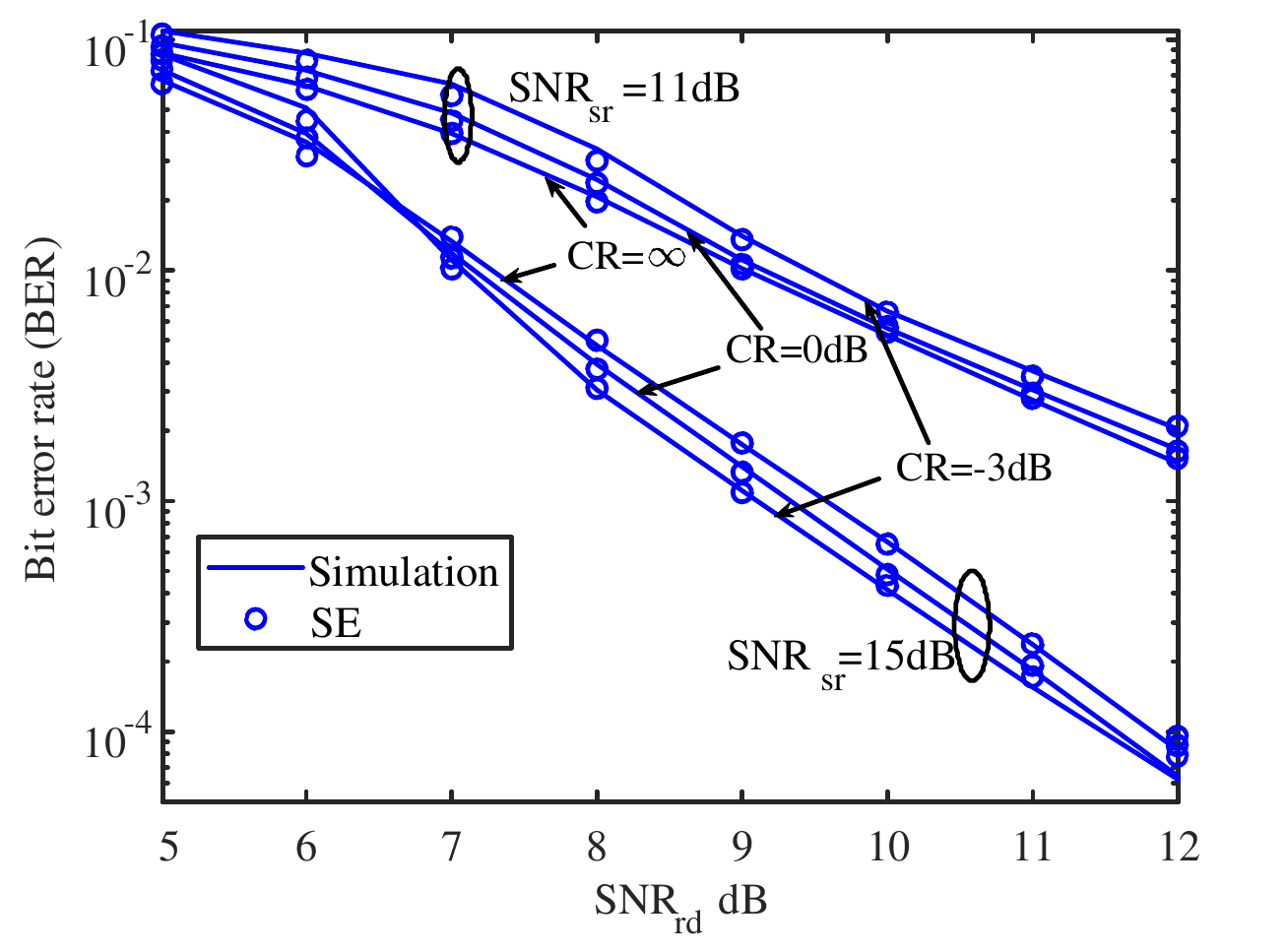}\\\vspace{-0.15cm}
  \caption{Simulated and predicted MSE for different clipping ratios $CR = [ -3, 0, 3, \infty]$. The other parameters are the same as these in Fig. \ref{Fig:Qant_net}.}\label{Fig:clips_net} 
\end{figure}

\subsection{Numerical Comparisons for SMV-MT}
We first examine the simple scenario where there is only one data stream in the source node, i.e., $M=1$. In this case, ${\bm X}_s$ is a vector with BPSK modulation and the system in \eqref{Eqn:mesh_CQ} can be modelled as an SMV-MT model. 

Fig. \ref{Fig:Qant_net} gives the simulated bit error rate
(BER) for GIP-MT in \eqref{Eqn:Ite_GIPMC} (i.e., without using GSO), methods 1-3 in \eqref{Eqn:method1}-\eqref{Eqn:method3}, as well as A-OAMP. (When $M=1$, A-OAMP can be equivalently called OAMP-MT.) We can clearly see the superior performance advantage of A-OAMP over other alternatives.  
 
 Fig. \ref{Fig:clips_net} gives the BERs of A-OAMP for systems with different clipping ratios. We can see that SE matches well with simulation in all cases. It is interesting to see that clipping actually leads to improved performance at $SNR_{sr}=15dB$. This counter-intuitive phenomenon was first reported in \cite{Liang2019}. This observation can be explained using a matching principle originally developed for the optimization of low-density parity-check codes. For more details, see \cite{Liang2019}.
 
\subsection{Numerical Comparisons for MMV-MT}
We next  consider $M = 2$, in which case the system in \eqref{Eqn:mesh_CQ} can be modelled as MMV-MT. Each data stream in ${\bm X}_s$ is BPSK modulated before being transmitted in two consecutive time slots. The symbols in these two streams are IID and are indexed as follows:
\BS
\begin{align}
{{\bm x}_{1,s}} = [ {{x_{1,1}}, \cdots ,{x_{1,{n_s}}}, \cdots ,{x_{1,{N_s}}}} ], \\
{{\bm x}_{2,s}} = [ {{x_{2,1}}, \cdots ,{x_{2,{n_s}}}, \cdots ,{x_{2,{N_s}}}} ],
\end{align}
\ES
where $x_{i,n_s}=-1$ or 1, $i=1,2$, $n_s=1,2,\cdots,N_s$. We assume that the two corresponding symbols in the two streams with the same index are correlated with the symmetric correlation index defined below:
\BS
\begin{align}
\Pr \big\{ { {{x_{2,{n_s}}} = -1} |{x_{1,{n_s}}} = 1} \big\} = \alpha, \\
\Pr \big\{ {{{x_{2,{n_s}}} = 1} |{x_{1,{n_s}}} =-1} \big\} = \alpha ,
\end{align}
\ES
with $\alpha$ the transition probability. 

The above system may find applications in a system with correlated sources. For example, a sensor may generate 2 data streams, one for temperature and one for humility.
\begin{figure}[t]
  \centering
  \includegraphics[width=0.35\textwidth]{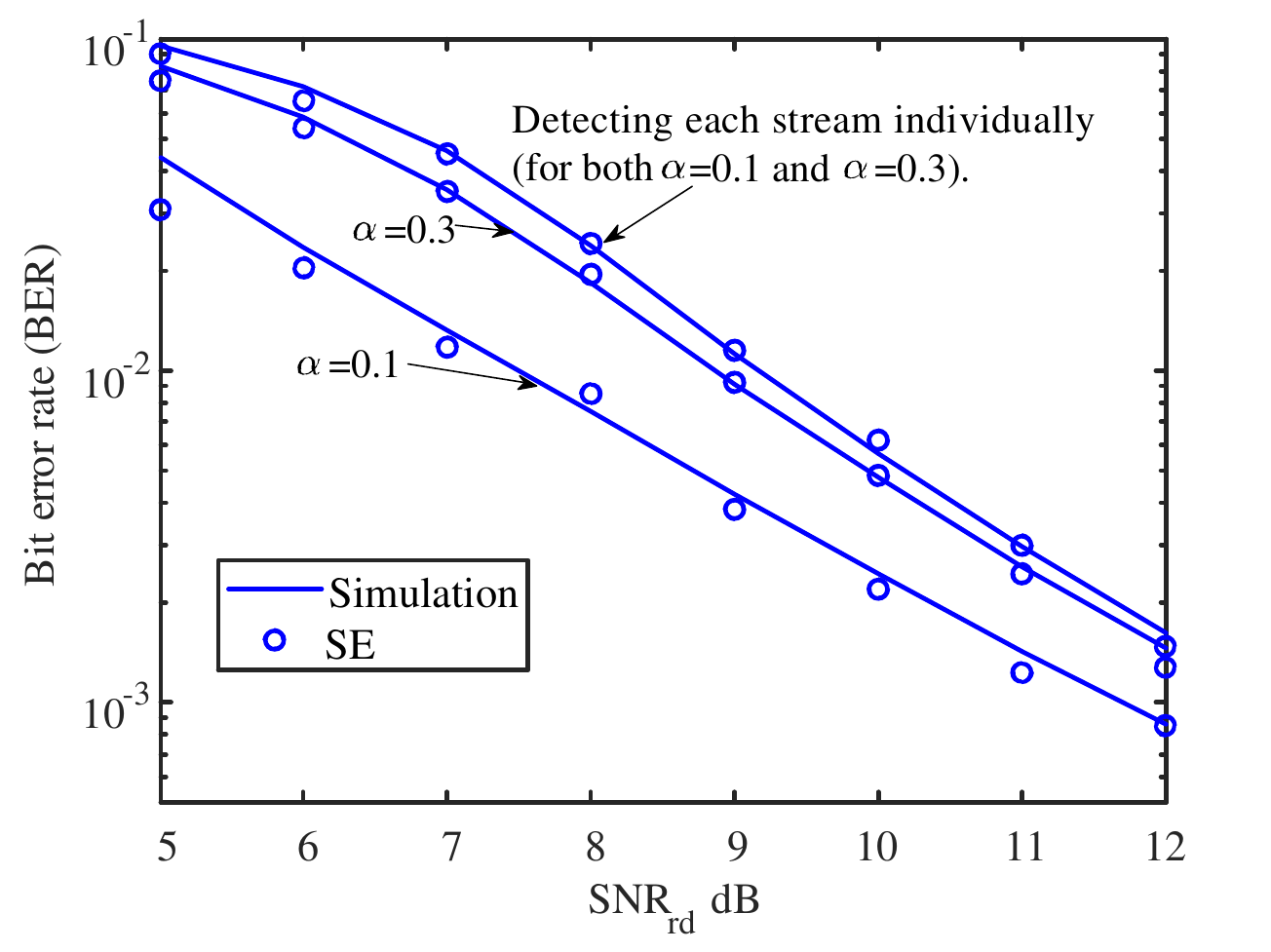}\\ \vspace{-0.12cm}
  \caption{Simulated and predicted MSE for different transmission probability $\alpha$. $M=2$. The other parameters are the same as these in Fig. \ref{Fig:Qant_net}.}\label{Fig:Rates_net}
\end{figure} 

Fig. \ref{Fig:Rates_net} shows the BERs for the above system, in which A-OAMP is used to detect the two streams jointly by treating them as two columns in an MMV problem. As a comparison, we also provide the simulation results obtained by detecting each stream individually. We can see that A-OAMP provides significantly improved performance by exploiting the correlation at the source. SE and simulation results also match well with each other.

In this paper, we focus on the effectiveness of A-OAMP. We paid less attraction to the application details in the simulation examples. We believe that A-OAMP can find useful applications in practice. For example, in medical testing (e.g., Covid-19), there can be multiple indicators, such as PCR level, antibody and lateral flow (LF) \cite{Lee2020,Alh2022}. These indicators can be correlated. Recently, compressed-sensing and AMP have been considered in massive-scale population screening based on such testing \cite{Iru2021}. A-OAMP can be applied to increase testing accuracy in such problems.

\section{Conclusion}
We extended OAMP from the basic SMV-ST model to more general MMV and MT ones and proved the related SE recursions. We presented an application in an MIMO relay system. The existing methods have difficulties in handling such non-ideal effects. The A-OAMP algorithm presented in this paper offers an efficient solution with excellent performance. 

MMV and MT problems represent two extremes, the former with identical $\{\bf{V}_k\}$ and the latter with independent $\{\bf{V}_k\}$. More general cases with partially correlated $\{\bf{V}_k\}$ are open problems for future research.
	

\appendices  
\section{Constrained Haar Transform}\label{APP:ConstrainedHaar}
For ${\bm V} \sim {{\bm {\mathcal H}}^{N \times N}}$, we call $\bf {VF}$ the Haar transform 
of $\bf F$. We will derive the properties of Haar transform under a linear constraint. Our derivations are based on a conjecture that we are not able to prove yet.
\subsection{A Basic Property of Haar Transform}
\begin{lemma} \label{Lem:Haar_CJGRJG}
Let ${\bm V}\sim \bm {\cal{H}}^{N \times N}$ conditional on any fixed ${\bm E} \in \mathbb{R}^{L \times N}$ and ${\bm F} \in \mathbb{R}^{N \times M}$, with $N \to \infty$ and $M$ finite. Then we have (i) ${\bm E}{\bm V}{\bm F}$ converges to both column-wise joint Gaussian (CJG) and row-wise joint Gaussian (RJG) in general; (ii) ${\bm E}{\bm V}{\bm F}$ converges to CPIIDG-RJG if ${\bm E}$ is orthogonal (i.e.,${\bm E} \in \bm {\mathcal U}^{N \times N}$); and (iii) ${\bm V}{\bm F}$ converges to CPIIDG-RJG.
\end{lemma}
\begin{IEEEproof}
We focus on ${\bm E}_{-1}$, the first row of ${\bm E}$. We construct ${\bm \Omega} \in \bm {\mathcal U}^{N \times N}$ with ${{\boldsymbol \Omega}_{ - 1}} = {{\bm E}_{ - 1}}/\| {{{\bm E}_{ - 1}}} \|$. (Note that we do not have any requirement on ${{\bm \Omega}_{ - n}}$ for $n > 1$.) From Lemma~\ref{lem:Haar}, ${\bm \Omega}{\bm V} \sim \bm{\cal{H}}^{N \times N}$, and so from Lemma~\ref{Lem:generalV}, for $\forall {\bm c} \in \mathbb{R}^{M}$, ${\bm \Omega}{\bm V}{\bm F}{\bm c}$ converges to PIIDG in distribution. Since any linear combinations of the entries in ${{\bm \Omega}_{ - 1}}{\bm V}{\bm F}$ is Gaussian, ${{\bm \Omega}_{ - 1}}{\bm V}{\bm F}$ is joint Gaussian, and thus the first row of ${\bm E}{\bm V}{\bm F}$ is joint Gaussian [Definition 3.2.5, \cite{Tong1990}]. For $l > 1$, we can prove that ${{\bm E}_{ - l}}{\bm V}{\bm F} $ is joint Gaussian in the same way by letting ${{\bm \Omega}_{ - 1}} = {{\bm E}_{ - l}}/\| {{{\bm E}_{ - l}}} \|$, so the matrix ${\bm E}{\bm V}{\bm F}$ converges to RJG. Similarly, we can prove $({\bm E}{\bm V}{\bm F})^{\rm T}$ is RJG, so ${\bm E}{\bm V}{\bm F}$ converges to both CJG and RJG. Hence claim (i) holds. When ${\bm E}$ is orthogonal (i.e., ${\bm E} \in \bm {\mathcal U}^{N \times N}$), ${\bm E}{\bm V} \sim {\bm {\cal H}}^{N \times N}$, so each column of ${\bm E}{\bm V}{\bm F}$ converges to PIIDG from Lemma~\ref{Lem:generalV}, and thus ${\bm E}{\bm V}{\bm F}$ converges to CPIIDG-RJG from (i). Hence claim (ii) holds for orthogonal ${\bm E}$. Finally, claim (iii) is a special case of (ii) when ${\bm E}={\bm I}^N$.
\end{IEEEproof}

Lemma~\ref{Lem:Haar_CJGRJG} shows that the entries of ${\bm {EVF}}$ converge to joint Gaussian. This result is useful below.

\subsection{Composite Structure}
We now consider a composite matrix structure:
\BE \label{Eqn:APP:compositeG}
{\bm G} = {\boldsymbol \Omega}\left[ {\begin{array}{*{20}{c}}
{\boldsymbol \varDelta} \\
{{\bm {VF}}}
\end{array}} \right] = {{\boldsymbol \Omega}^{\parallel}}{\boldsymbol \varDelta}  + {{\boldsymbol \Omega}^ \perp }{\bm {VF}}, 
\EE
where ${\boldsymbol \Omega}=[{\boldsymbol \Omega}^{\parallel},{\boldsymbol \Omega}^{\perp}] \in {\boldsymbol {\cal{U}}}^{N \times N}$, ${\boldsymbol \varDelta} \in \mathbb{R}^{J \times M}$, ${\bm V}\sim \boldsymbol{\cal{H}}^{(N-J) \times (N-J)}$ and ${\bm F} \in \mathbb{R}^{(N-J) \times M}$. We will always assume that $N \to \infty$ and $J$ remains finite below. In this case, from Lemma~\ref{Lem:Haar_CJGRJG}, ${\bm {VF}}$ converges to CPIIDG-RJG, so ${\bm {VF}}$ has identical row distribution. We construct ${\boldsymbol \varDelta}$ as follows.
\begin{itemize}
\item Each row of ${\boldsymbol \varDelta}$ is independently drawn from the row distribution of ${\bm {VF}}$, and
\item ${\boldsymbol \varDelta}$ is independent of ${\bm {VF}}$.
\end{itemize}

We first prove a useful result. Let ${\boldsymbol \Omega} \in {\boldsymbol {\cal U}}^{N \times N}$ be a random orthogonal matrix (not necessarily Haar) and ${\bm a}$ be PIIDG conditional on ${\boldsymbol \Omega}$ with entry-wise distribution $\mathcal{N}(\mu,v)$.

\begin{lemma} \label{Lem:UEuncorrelated}
${\boldsymbol \Omega}{\bm a}$ converges to PIIDG, and each entry of ${\boldsymbol \Omega}{\bm a}$ is asymptotically independent of ${\boldsymbol \Omega}$.
\end{lemma}
\begin{IEEEproof}
The entries of ${\boldsymbol \Omega}{\bm a}$ are Gaussian (since ${\bm a}$ is PIIDG). Besides, the covariance of ${\boldsymbol \Omega}{\bm a}$ is,
\BE
{\rm {Cov}}({\boldsymbol \Omega}{\bm a}, {\boldsymbol \Omega}{\bm a})={\boldsymbol \Omega}{\rm Cov}({\bm a}, {\bm a}){\boldsymbol \Omega}^{\rm T}=(\|\bm a\|^2/N)\cdot {\bm I}^N,
\EE
which shows that ${\boldsymbol \Omega}{\bm a}$ is entry-wise uncorrelated. Thus ${\boldsymbol \Omega}\!{\bm a}$ is PIIDG. Its entry-wise distribution is $\mathcal{N} (\mu,v)$, which is regardless of the selection of ${\boldsymbol \Omega}$ (provided that ${\boldsymbol \Omega}$ is orthogonal).
\end{IEEEproof}

We next derive a few properties of ${\bm G}$ in \eqref{Eqn:APP:compositeG}.
\begin{lemma}  \label{Lem:APP:structureG}
In \eqref{Eqn:APP:compositeG}, ${\bm G}$ converges to CPIIDG-RJG and each entry of ${\bm G}$ is asymptotically independent of ${\boldsymbol \Omega}$.
\end{lemma}
\begin{IEEEproof}
From Lemma~\ref{Lem:Haar_CJGRJG}, ${\bm {VF}}$ converges to CPIIDG-RJG, so $\left[\!\!\!\!{\begin{array}{*{20}{c}}{\boldsymbol \varDelta} \\{{\bm {VF}}}\end{array}} \!\!\!\!\right]$ is also CPIIDG-RJG from the construction of ${\boldsymbol \varDelta}$. Then, using Lemma~\ref{Lem:UEuncorrelated}, we see that ${\bm G}$ converges to CPIIDG-RJG and is asymptotically entry-wise independent of ${\boldsymbol \Omega}$.
\end{IEEEproof}

We now add an additional condition that ${\boldsymbol \Omega}^{\parallel}$ is CPIID in \eqref{Eqn:APP:compositeG}. As an example, ${\boldsymbol \Omega}^{\parallel}$ can be a base of a CPIID matrix $\bm B$. Such ${\boldsymbol \Omega}^{\parallel}$ is CPIID as well as column-orthonormal.

\begin{lemma} \label{Lem:APP:structureGperp}
Under the same conditions as Lemma~\ref{Lem:APP:structureG} plus ${\boldsymbol \Omega}^{\parallel}$ being CPIID, ${\boldsymbol \Omega}^{\perp}{\bm {VF}}$ converges to CPIIDG-RJG and each entry of ${\boldsymbol \Omega}^{\perp}{\bm {VF}}$ is asymptotically independent of ${\boldsymbol \Omega}$ in \eqref{Eqn:APP:compositeG}.
\end{lemma}
\begin{IEEEproof}
The $m^{\rm{th}}$ column of ${\boldsymbol \Omega}^{\perp}{\bm {VF}}$ is 
\BE   \label{Eqn:APP:columnvf}
{\boldsymbol \Omega}^{\perp}{\bm {VF}}_{|m} = {\bm G}_{|m} - {\boldsymbol \Omega}^{\parallel}{\boldsymbol \varDelta}_{|m}. 
\EE
From Lemma~\ref{Lem:APP:structureG} and since ${\boldsymbol \Omega}^{\parallel}$ is CPIID, both ${\bm G}_{|m}$ and ${\boldsymbol \Omega}^{\parallel}{\boldsymbol \varDelta}_{|m}$ are PIID. (Note that ${\boldsymbol \Omega}^{\parallel}{\boldsymbol \varDelta}_{|m}$ may not be Gaussian.) Therefore,  when $ N \to \infty $ and $J$ remains finite,
\BE \label{Eqn:approximate}
\frac{{{\rm{E}}\{ {{{ \big|{{{[ {{\boldsymbol \Omega}^{\parallel} {\boldsymbol \varDelta}} ]_{n,m}}}} \big|}^2}} \}}}{{{\rm{E}}\{ {{\big|{ {{[{\bm G}]_{n,m}}} \big|}^2}} \}}}=\frac{{{{\| {{{\boldsymbol \Omega}^\parallel}{\boldsymbol \varDelta} }_{|m} \|}^2}}}{{{{\|\bm G_{|m} \|}^2}}} = \frac{J}{N} \to 0,
\EE
where the subscript ``$n,m$'' indicates the entry in the $n^{\rm{th}}$ row and $m^{\rm{th}}$ column. Hence, in \eqref{Eqn:APP:columnvf}, ${\boldsymbol \Omega}^{\parallel}{\boldsymbol \varDelta}_{|m}$ is negligible relative to ${\bm G}_{|m}$, so ${{\boldsymbol \Omega}^\perp}{\bm {VF}}$ converges to $\bm G$ in distribution. Then Lemma~\ref{Lem:APP:structureGperp} follows Lemma~\ref{Lem:APP:structureG}.
\end{IEEEproof}

\subsection{Constrained Haar Distribution}
Start from a special case. Let $N > J$ and denote
\BE \label{Chap2:Eqn:defineJ}
   {{\bm J}^{N \times J}} = \left[ {\begin{array}{*{20}{c}}
{{{\bm I}^{J \times J}}}\\
{{{\bm 0}^{( {N - J} ) \times J}}}
\end{array}} \right].
\EE
Define a distribution, denoted as ${{\bm V}^\# } \!\!\!\sim \!\bm {\mathcal H}( {{{\bm J}^{N \times J}} \!\!=\! {{\bm V}^\# }{{\bm J}^{N \times J}}}) $, if ${{\bm V}^\# }$ is uniformly distributed over the sample space of $\bm{\mathcal{H}}^{N \times N}$ under a linear constraint
\BE \label{Chap2:Eqn:constraintJ}
{\bm J}^{N \times J} = {{\bm V}^\# }{{\bm J}^{N \times J}}.
\EE
We can show that such ${{\bm V}^\# }$ must have the following structure
\BE \label{Chap2:Eqn:structureV}
{{\bm V}^\# } = \left[ {\begin{array}{*{20}{c}}
{{{\bm I}^{J \times J}}}&{{{\bm 0}^{J \times ( {N - J} )}}}\\
{{{\bm 0}^{{(N-J)} \times  J }}}&{\tilde {\bm V}}
\end{array}} \right],
\EE
where $\tilde {\bm V} \sim {\bm{\mathcal H}^{( {N - J}) \times (N-J) }}$.

Now consider a more general case. For fixed $\bm A$, ${\bm B} \in \mathbb{R}^{N \times J}$, denote ${\bm V} \sim \bm{\mathcal H}( {{\bm A} = {{\bm V}{\bm B}}})$ if $\bm V$ is uniformly distributed over the subset of the sample space of ${\bm{\mathcal H}^{N \times N}}$ under a constraint:
\BE \label{Chap2:Eqn:A=VB_Appendix}
    {\bm A}={\bm V}{\bm B}.
\EE
The above is a special case of Bolthausen’s conditioning problem \cite{Bayati2011,Takeuchi2020,Bolthausen2014}. We will assume that ${\rm {rank}}(\bm B)\!\!=\!\!J$. Define an orthogonal matrix 
\BE \label{Eqn:APP:W}
{\bm W}=[{\bm W}^\parallel, {\bm W}^\perp],
\EE
where ${\bm W}^\parallel \in \mathbb{R}^{N \times J}$ and ${\bm W}^\perp \in \mathbb{R}^{ N \times (N-J)}$ form the bases of ${\rm{Sp}}( \bm B )$ and ${\rm{Sp}}(\bm B)^\perp$, respectively. (Note: The sizes of ${\bm W}^\parallel$ and ${\bm W}^\perp$ should be adjusted if ${{\rm {rank}}(\bm B)}<J$, but all the results below still hold.) Let ${\bm V}^S$ be an arbitrary sample in ${\bm V}\! \sim\! \bm{\mathcal H}( {{\bm A}\! = \!{{\bm V}{\bm B}}})$ such that
\BE \label{Chap2:Eqn:VSsample}
{\bm V}{\bm B}= {\bm V}^S{\bm B}.
\EE
Then $\bm V$ above is connected to ${\bm V}^\#$ in \eqref{Chap2:Eqn:structureV} as follows:
\BE  \label{Chap2:VQ=VSWWQ}
{\bm V}{\bm B}\overset{(a)}={\bm V}^S{\bm B}\overset{(b)}={\bm V}^S{\bm W}{\bm W}^{\rm T}{\bm B}\overset{(c)}={\bm V}^S{\bm W}{\bm V}^{\#}{\bm W}^{\rm T} {\bm B},
\EE
where (a) follows \eqref{Chap2:Eqn:VSsample}, (b) holds as ${\bm W}$ is orthogonal and (c) holds since ${{\bm W}^{\rm{T}}}{\bm B} = {{\bm J}^{N \times J }}{( {\bm W}^\parallel )^{\rm{T}}}{\bm B}$ and so, from \eqref{Chap2:Eqn:constraintJ}, ${\bm V}^{\#}{\bm W}^{\rm T}{\bm B}={\bm W}^{\rm T}{\bm B}$. From \eqref{Chap2:VQ=VSWWQ}, for ${\bm V} \sim \bm{\mathcal H}( {{\bm A} = {{\bm V}{\bm B}}})$, we can write
\BE \label{Chap2:Eqn:denotenewV}
    {\bm V} ={{\bm V}^S}{\bm W}{{\bm V}^\# }{{\bm W}^{\rm{T}}}. 
\EE
\begin{lemma}
With $\tilde {\bm V}$ defined in \eqref{Chap2:Eqn:structureV} and some ${\bm C} \in \mathbb{R}^{J \times M}$, we can write
\BS
\begin{eqnarray}
&{\bm {VF}}={\bm {AC}}, \quad\quad\quad\quad\quad\;\;\;\;\; {\text{ if }} {\rm{Sp}}( \bm F ) \subseteq {\rm{Sp}}(\bm B ), \label{Eqn:Chap2:parrelVF}\\
&{\bm {VF}}={{\bm V}^S}{\bm W}^{\perp}{\tilde {\bm V}{{( {\bm W}^\perp)}^{\rm{T}}}{\bm F}},{\text{ if }} {\rm{Sp}}( \bm F ) \perp {\rm{Sp}}(\bm B). \label{Eqn:Chap2:perpVF}
\end{eqnarray}
\ES
\end{lemma}
\begin{IEEEproof}
If ${\rm{Sp}}( \bm F ) \!\subseteq\! {\rm{Sp}}(\bm B )$, we can write ${\bm F}\!=\!{\bm {BC}}$ with some $\bm C$, so \eqref{Eqn:Chap2:parrelVF} holds since ${\bm A}\!=\!{\bm V}{\bm B}$. Combining \eqref{Chap2:Eqn:structureV}, \eqref{Eqn:APP:W} and \eqref{Chap2:Eqn:denotenewV} leads to \eqref{Eqn:Chap2:perpVF} since $({\bm W}\!^\parallel\!)^{\rm T}{\bm F}\!=\!{\bm 0}^{J \!\times\! M}$ in this case.
\end{IEEEproof}

We now consider a Markov chain
\BE \label{Eqn:Markov_S}
{\bm z} \to [{\bm F},{{\bm A},{\bm B}}] \to{\bm V} \sim {\boldsymbol {\mathcal H}}({ {{\bm A} = {\bm V}{\bm B}} }).
\EE
Then, the distribution of $\bm V$ is conditional on ${\bm F},{\bm A},{\bm B}$ and an arbitrary ${\bm z}$.
\begin{lemma}  \label{Lem:VF_perp}
Let $N \to \infty$, $J$ remain finite, $\bm A$ be CPIID and ${\rm{Sp}}( \bm F ) \perp {\rm{Sp}}(\bm B )$. Then, in \eqref{Eqn:Chap2:perpVF}, (i) ${\bm {VF}}$ converges to CPIIDG-RJG and (ii) ${\bm {Vf}}/||\bm f||$ is asymptotically entry-wise independent of $\bm z$, $\forall {\bm f} \in {\rm{Clmn}}(\bm F)$.
\end{lemma}
\begin{IEEEproof}
Denote ${\boldsymbol \Omega}^{\parallel}\equiv{\bm V}^S{\bm W}^{\parallel}$ and ${\boldsymbol \Omega}^{\perp}\equiv{\bm V}^S{\bm W}^{\perp}$. Clearly, ${\boldsymbol \Omega}=[{\boldsymbol \Omega}^{\parallel}, {\boldsymbol \Omega}^{\perp}]$ is orthogonal. Then \eqref{Eqn:Chap2:perpVF} can be rewritten as
\BE  \label{Eqn:Chap2:VFU}
{\bm {VF}}={\boldsymbol \Omega}^{\perp}{\tilde {\bm V}}({\bm W}^{\perp})^{\rm T}{\bm F}.
\EE
Recall that ${\bm W}^{\parallel}$ is a base of ${\rm {Sp}}(\bm B)$ and ${\bm V}^S$ is a sample of ${\bm V} \sim {\boldsymbol {\cal{H}}}({{\bm A}={\bm {VB}}})$. Hence ${\boldsymbol \Omega}^{\parallel}  \subseteq {\rm{Sp}}(\bm A)$ and so ${\boldsymbol \Omega}^{\parallel}$ is CPIID. Then from Lemma~\ref{Lem:APP:structureGperp}, ${\bm {VF}}$ in \eqref{Eqn:Chap2:VFU} converges to CPIIDG-RJG and is entry-wise independent of ${\boldsymbol \Omega}^{\perp}$ asymptotically, so (i) holds. For (ii), note that \eqref{Eqn:Chap2:VFU} involves ${\bm F}$ explicitly, and $\bm A$ and $\bm B$ implicitly since ${\boldsymbol \Omega}^{\perp}$ and ${\bm W}^{\perp}$ are the bases of ${\rm{Sp}}( \bm A )^{\perp}$ and ${\rm{Sp}}( \bm B)^{\perp}$ respectively. We address this issue as follows.
\begin{itemize}
\item From Lemma~\ref{Lem:APP:structureGperp}, ${\bm {VF}}$ in \eqref{Eqn:Chap2:VFU} is asymptotically entry-wise independent of $\bm A$, since ${\boldsymbol \Omega}^{\parallel}  \subseteq {\rm{Sp}}(\bm A)$ and ${\boldsymbol \Omega}^{\perp}  \subseteq {\rm{Sp}}(\bm A)^{\perp}$.
\item From \eqref{Eqn:Chap2:VFU}, we write ${\bm {Vf}}\!/\|\bm f\|\!=\!{\boldsymbol \Omega}^{\perp}{\tilde {\bm V}}({\bm W}^{\perp})^{\rm T}{\bm f}\!/\|\bm f\|$, in which ${\tilde {\bm V}}({\bm W}^{\perp})^{\rm T}{\bm f}\!/\|\bm f\|$ is uniformly distributed on a unit sphere from Lemma~\ref{lem:Haar}, which is independent of $({\bm F},{\bm A},{\bm B})$.
\end{itemize}
Combing the above, we can see that (ii) holds.
\end{IEEEproof}

\subsection{General Constrained Haar Transform}
We now derive an extension of Lemma \ref{Lem:VF_perp}. Consider partitions: ${\bm A}=[{\bm A}_*,{\bm B}_{**}]$ and ${\bm B}=[{\bm B}_*,{\bm B}_{**}]$ with ${\bm A}_*={\bm V}{\bm B}_*$ and ${\bm A}_{**}={\bm V}{\bm B}_{**}$. 
\begin{assumption}\label{Asu:generalHaar}
(a) ${\bm A}_{**}$ converges to CPIIDG-RJG and ${\bm a}/\|\bm a\|, \forall {\bm a} \in {\rm{Clmn}}({\bm A}_{**})$, is entry-wise independent of $\bm z$ in \eqref{Eqn:Markov_S}; (b) ${\rm{Sp}}( \bm F )\perp {\rm{Sp}}( {{{\bm B}_*}} )$ and (c) ${\bm A}=[{\bm A}_*,{\bm A}_{**}]$ is CPIID. 
\end{assumption}

Then we have the following lemma.
\begin{lemma} \label{Chap2:Lem:P=VQ_perp}
Under Assumption \ref{Asu:generalHaar} and letting $N \to \infty$, $J$ fixed, we have (i) $[{\bm A}_{**}, {\bm V}{\bm F}]$ converges to CPIIDG-RJG in distribution, and (ii) ${\bm V}{\bm f}/\|{\bm f}\|, \forall {\bm f} \in {\rm {Clmn}}(\bm F)$ is entry-wise independent of $\bm z$ in \eqref{Eqn:Markov_S}.
\end{lemma}
\begin{IEEEproof}
We can always decompose ${\bm F}={\bm F}'+{\bm F}{''}$, where ${\rm{Sp}}({\bm F}') \!\perp\! {\rm{Sp}}({\bm B}_{**})$ and ${\rm{Sp}}({\bm F}{''}) \subseteq {\rm{Sp}}({\bm B}_{**})$. Consider two special cases of either ${\bm F}'$ or ${\bm F}{''}$ being empty.
\begin{itemize}
\item First, let ${\bm F}'$ be empty. In this case, ${\rm{Sp}}(\bm F) \!\subseteq\! {\rm{Sp}}({\bm B}_{**})$, so ${\rm{Sp}}({\bm V}{\bm F}) \!\subseteq \! {\rm{Sp}}({\bm A}_{**})$ from \eqref{Eqn:Chap2:parrelVF}. Then claims (i) and (ii) hold directly from condition~(a) (i.e., Assumption \ref{Asu:generalHaar}(a)).
\item Next, let ${\bm F}{''}$ be empty. In this case, ${\rm{Sp}}({\bm F}) \!\!\perp\! {\rm{Sp}}({\bm B})$. Then claims (i) and (ii) follow Lemma~\ref{Lem:VF_perp} under condition (c).
\end{itemize}
Combining these two cases, we can show claims (i) and (ii) hold for ${\bm F}\!=\!{\bm F}'\!+\!{\bm F}{''}$.
\end{IEEEproof}

\underline{{\textbf{Notes:}}} For convenience, we call $\{ {{\bm f}/\| {\bm f} \|,\forall {\bm f} \in {\rm{Clmn}}(\bm F)} \}$ the angles of $\bm F$. We say a matrix is ``desirable'' if it is CPIID-RJG and its angles are entry-wise independent of any $\bm z$. ${\bm A}_*$ above is assumed to be CPIID, but it can be non-Gaussian and correlated with $\bm z$, so it is not necessarily ``desirable''.

From Lemma \ref{Lem:generalV}, $\bm {Vf}$ is desirable if $\bm V$ is fully Haar distributed. The situation is more complicated for ${\bm V} \!\sim\! {\bm{\mathcal H}}( {{\bm A} \!= \!{\bm {VB}}})$ when ${\bm A}_*$ above, is not necessarily desirable, as ${\rm{Sp}}( \bm {VF} )$ may fall into ${\rm{Sp}}( \bm {A_*} )$ and become ``undesirable''. Lemma \ref{Chap2:Lem:P=VQ_perp} says that such situation can be avoided if we can ensure ${\rm{Sp}}( \bm {VF} )\perp {\rm{Sp}}( \bm {A_*} )$. This orthogonal principle is the basis of OAMP-MMV and the related algorithms. Also, note that the result is approximate due to \eqref{Eqn:approximate}.


\end{document}